\documentclass[twocolumn]{aastex701}

\usepackage{subfig}
\usepackage{placeins}
\newcommand{\redbf}[1]{#1}

\begin{document}

\title{The Hand of God Revisited: Revealing the Shell Structure and Jet–Supernova Remnant Interaction in MSH 15--52}

\author[orcid=0009-0006-1753-7623,sname='Dang']{Lingxiao Dang}
\affiliation{School of Astronomy $\&$ Space Science, Nanjing University, 163 Xianlin Avenue, Nanjing 210023, People's Republic of China}
\affiliation{Department of Physics \& Astronomy, University of Manitoba, Winnipeg, MB R3T~2N2, Canada}
\email[show]{dangl@myumanitoba.ca}

\author[orcid=0000-0002-4400-4703,sname='Sander']{Isabel Sander}
\affiliation{Department of Physics \& Astronomy, University of Manitoba, Winnipeg, MB R3T~2N2, Canada}
\email{sanderi@myumanitoba.ca}  

\author[orcid=0000-0001-6189-7665,sname='Safi-Harb']{Samar Safi-Harb}
\affiliation{Department of Physics \& Astronomy, University of Manitoba, Winnipeg, MB R3T~2N2, Canada}
\email[show]{samar.safi-harb@umanitoba.ca}  

\author[orcid=0000-0003-2624-0056,sname='Fryer']{Christopher Fryer}
\affiliation{Center for Nonlinear Studies, Los Alamos National Laboratory, Los Alamos, NM 87545, USA}
\affiliation{Department of Physics, The George Washington University, Washington, DC 20052, USA}
\email{fryer@lanl.gov}

\begin{abstract}

We present an X-ray analysis of the supernova remnant (SNR) 
MSH~15--52 associated with the young rotation-powered pulsar PSR~B1509--58, using \textit{Chandra}, \textit{XMM-Newton}, and \textit{eROSITA}, complemented by \textit{MeerKAT} radio data.
We report the discovery of the missing southern counterpart to the bright northern RCW~89 region and identify
a foreground X-ray source that may be a previously unidentified SNR candidate. Imaging and spectroscopy with \textit{XMM-Newton} and \textit{eROSITA} reveal the shell properties of MSH~15--52 and suggest that both the southern and northern shell-like structures are consistent with shocked interstellar medium associated with the forward shock, while their low ionization timescales and possible hard power-law components indicate possible influence from particle injection or additional heating by the pulsar wind nebula (PWN) jets. New \textit{Chandra} spectral maps of the southern jet reveal a region of hard non-thermal emission, with a photon index of $\sim$1.6, suggesting interaction between the PWN and the reverse shock and possible re-acceleration of the pulsar outflow.
Spatially resolved spectroscopy of the RCW 89 northern clumps reveals high-velocity, metal-rich ejecta with enhanced Ne and Mg abundances. Comparison with core-collapse nucleosynthesis models favors a massive progenitor with an initial mass of approximately 29~$M_\odot$
in one-dimensional yield models, while asymmetric explosion models show that lower-mass progenitors can also reproduce the observed abundance ratios.
Our study supports a picture in which RCW 89 is part of a larger asymmetric SNR shell interacting with energetic pulsar jets, making the system a valuable laboratory for probing pulsar--PWN--SNR interaction.


\end{abstract}

\section{Introduction}
SNRs serve as key laboratories for understanding the life cycles of massive stars and the physics of supernova (SN) explosions. 
They retain the nucleosynthesis products of their progenitors and play an essential role in enriching the interstellar medium (ISM) with heavy elements \citep{2017hsn..book.....A, 2018ApJ...856...63F}. X-ray observations of their hot plasma further allow us to probe shock dynamics, elemental abundances, and the interaction between the expanding ejecta and the surrounding ISM.

Various methods have been proposed to infer the progenitor properties of SNRs \citep{2009_lopez, 2014_yamaguchi}. 
With the advancement of SN nucleosynthesis models, abundance patterns derived from X-ray spectroscopy have become a powerful tool for constraining the progenitor mass and explosion characteristics \citep{Zhou_2019,2023_braun, 2026ApJ...997..211T, 2026ApJ..1003L...2B}.
 
MSH 15--52 (G320.4--1.2, also known as RCW~89) hosts a young and energetic pulsar \citep{1982_seward}, 
 PSR J1513--5908 (B1509--58)
 with an estimated characteristic age of 1.7 kyr \citep[][]{kaspi_1994,Seward_1983} and a distance of approximately 5.2 kpc \citep{Gaensler_1999}\footnote{see \url{http://snrcat.physics.umanitoba.ca/SNRrecord.php?id=G320.4m01.2} for a summary of its observations and properties}. 
Owing to its youth and brightness, it has been extensively studied in the radio \citep[][]{1961_Mills,Zhang_2025}, X-ray \citep[][]{1982_seward,2002_gaensler}, and gamma-ray bands \citep[][]{Aharonian_2005,2025_Egg}. Its X-ray morphology, 
popularly nicknamed the “Hand of God" because of its hand-like appearance in \textit{Chandra} images, further highlights the complexity of the system and has drawn broad attention to it.
The northern RCW~89 region is a thermally dominated complex
whose X-ray emission is believed to be shaped by its interaction with the PWN to its south  (see Figure~\ref{fig:chandra_image}). Although RCW~89 is a prominent component
of the SNR MSH~15--52, the origin of its thermal X-ray emission remains debated. Some studies have argued that the emission is powered by a precessing jet emanating from the PWN \citep{Manchester_1987,Brazier_1997MNRAS,Seward_1983}. \citet{Yatsu_2005} analyzed seven bright knots in RCW~89 and reported a clockwise sequence of temperature and ionization timescale variations, which they interpreted as evidence for jet precession. In contrast,  \citet{Borkowski_2020} identified a possible blast-wave structure in \textit{Chandra} images, suggesting that at least part of the emission may instead arise from the SNR shock.

Furthermore, although the PWN and RCW~89 structure in MSH 15--52 
have been extensively studied, the presumably extended SNR shell has received far less attention, largely because of limited photon statistics, the restricted fields of view of previous X-ray observations, and the focus on the PWN and RCW~89 complex. We note, however, that an early study with \textit{ROSAT} \citep{Trussoni_1996} reported a southern shell-like feature in soft X-rays, but the low count rate prevented a spectral characterization of this component. As a result, its spectrum was combined with that of the
bright PWN, leaving the physical nature of the southern shell-like structure largely unexplored.

Growing multi-wavelength evidence further points to a possible interaction between the PWN jet and the surrounding SNR shell. Such jet–shell interactions are uncommon among known SNRs, with W50 being one of the few prominent examples \citep{Safi-Harb_2022}. This potential coupling highlights MSH 15--52 as a particularly valuable laboratory for studying the interaction between an energetic PWN and its SNR environment.

In this work, we present a new X-ray analysis of the SNR~MSH 15--52 using observations from \textit{Chandra}, \textit{XMM-Newton}, and \textit{eROSITA}. These data provide complementary views of the system: the high angular resolution of \textit{Chandra} enables spatially resolved spectroscopy of the compact ejecta clumps in RCW~89 and the construction of new spectral maps of the PWN jet regions, while the sensitivity and wider fields of view of \textit{XMM-Newton} and \textit{eROSITA} allow us to trace and characterize the faint, extended shell emission.  We use the abundance patterns of the ejecta clumps to constrain the supernova progenitor properties and report the first identification and spectral characterization of a southern shell-like structure, likely the missing counterpart to the northern RCW~89 component. Finally, we compare 
the northern and southern shell properties and use the
\textit{Chandra} spectral maps to assess whether the pulsar outflow has impacted the surrounding SNR environment, providing evidence for a possible jet-shell interaction in the southeastern region of the system.

\begin{table*}[htbp]
\centering
\caption{Log of the X-ray observations used in this study.}
\begin{tabular}{lcccc}
\hline
\hline
Satellite & Date & ObsID & Pointing & Total Exposure \\
  &   &   &   RA, Dec (J2000) &  (ks) \\
\hline
\textit{Chandra} & 2000-08-14 & 754   & 15 13 55.60, -59 08 08.90 & 19.04 \\
\textit{Chandra} & 2003-10-18 & 3833  & 15 13 55.60, -59 08 09.20 & 19.13 \\
\textit{Chandra} & 2003-04-21 & 3834  & 15 13 55.60, -59 08 09.20 & 9.47  \\
\textit{Chandra} & 2003-04-28 & 4384  & 15 13 55.60, -59 08 09.20 & 9.89  \\
\textit{Chandra} & 2004-12-28 & 5534  & 15 13 55.60, -59 08 08.90 & 49.49 \\
\textit{Chandra} & 2004-12-31 & 5562  & 15 13 34.20, -59 01 58.00 & 29.58 \\
\textit{Chandra} & 2005-02-07 & 5535  & 15 13 55.60, -59 08 08.90 & 42.59 \\
\textit{Chandra} & 2005-04-29 & 6116  & 15 13 55.60, -59 08 08.90 & 47.04 \\
\textit{Chandra} & 2005-10-18 & 6117  & 15 13 55.60, -59 08 08.90 & 45.55 \\
\textit{Chandra} & 2008-06-19 & 9138  & 15 13 34.20, -59 01 58.00 & 58.98 \\
\textit{Chandra} & 2017-02-02 & 18023 & 15 13 55.50, -59 08 08.80 & 18.32 \\
\textit{Chandra} & 2017-12-27 & 20910 & 15 13 37.27, -59 01 46.20 & 46.50 \\
\textit{Chandra} & 2018-01-05 & 19299 & 15 13 37.27, -59 01 46.20 & 43.54 \\
\textit{Chandra} & 2018-01-16 & 19300 & 15 13 37.27, -59 01 46.20 & 20.83 \\
\textit{Chandra} & 2018-02-12 & 20932 & 15 13 37.27, -59 01 46.20 & 30.70 \\
\textit{Chandra} & 2018-04-08 & 20933 & 15 13 37.27, -59 01 46.20 & 43.02 \\
\textit{Chandra} & 2022-09-20 & 23540 & 15 13 55.50, -59 08 08.80 & 10.65 \\
\textit{Chandra} & 2022-09-23 & 27448 & 15 13 55.50, -59 08 08.80 & 16.98 \\
\textit{XMM-Newton}  & 2005-08-07 & 0302730101  & 15 14 58.48, -59 14 08.0 & 26.81 \\
\textit{XMM-Newton}  & 2004-09-14 & 0207050101  & 15 14 58.48, -59 14 08.0 & 26.92 \\
\textit{eROSITA}  & 2019-09-10 & 900035  & 15 15 21.60, -59 06 00.00 & 20.58 \\
\textit{eROSITA}  & 2019-09-14 & 900043  & 15 15 24.00, -59 06 36.00 & 21.19 \\
\textit{eROSITA}  & 2020-02-27 & Early Data Release  & 15 02 51.40, -60 00 53.00 & 0.77 \\
\hline
\end{tabular}
\label{tab:xraylog}
\end{table*}

\section{Data}
\label{sec:data}

\subsection{\textit{Chandra}}

\begin{figure}[htbp]
    \centering
    \includegraphics[width=\columnwidth]{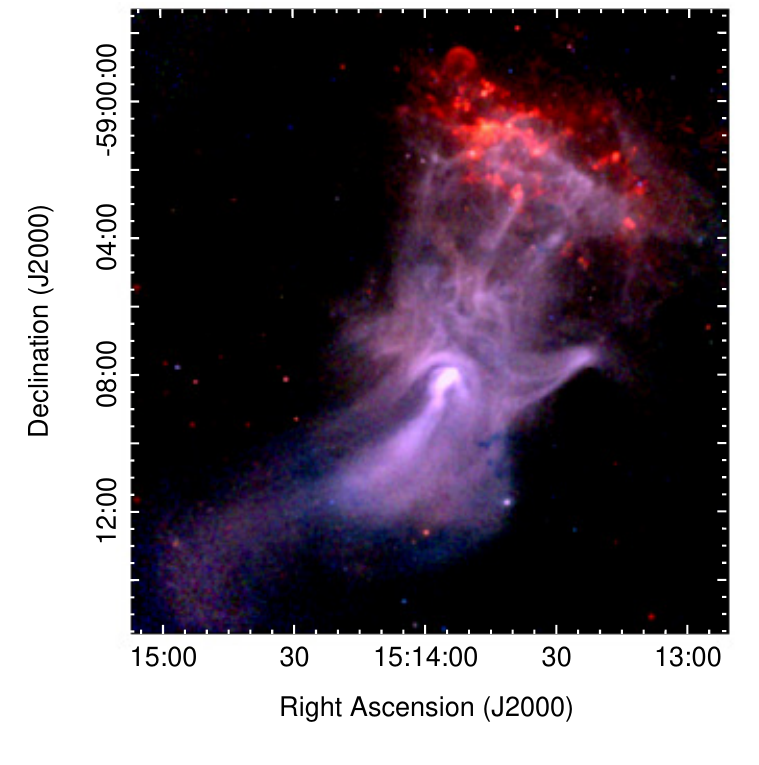}
    \caption{Tri-color \textit{Chandra} X-ray image of RCW 89 and its associated PWN: red: 500--2000 eV, green: 2000--3000 eV, blue: 3000--9000 eV. This image shows that the northern emission is dominated by soft X-rays from the RCW~89 thermal complex, while the prominent southern jet emanating from PSR~J1513--5908 (the brightest white knot near the image centre) is dominated by hard non-thermal X-ray emission.}
    \label{fig:chandra_image}
\end{figure}

\textit{Chandra} has observed MSH~15--52 many times from 2000 to 2022, as summarized in Table~\ref{tab:xraylog}. We used all available observations to obtain the image of the SNR shown in Figure \ref{fig:chandra_image}. For the spectral analysis, however, we restricted our study to the 2017--2018 observations, as the ejecta were reported to exhibit high velocities and temporal variability in several clumps \citep{Borkowski_2020}, and because these observations provide a deep view of the system with a total exposure of 184.59 ks. {The datasets used in this paper, obtained by the Chandra X-ray Observatory, are contained in the Chandra Data Collection ~\dataset[DOI: 10.25574/cdc.663]{https://doi.org/10.25574/cdc.663}.}

We processed the \textit{Chandra} data with the \textit{Chandra} Interactive Analysis of Observations (CIAO) 4.17 software package and CALDB 4.12.2 \citep{2006Fruscione}. The preprocessing was performed using \textit{chandra\_repro}, after which the individual observations were aligned with \textit{reproject\_obs}. Images were generated with \textit{flux\_obs}, and the source and background spectra were extracted using \textit{specextract}.

\begin{figure*}[htbp]
    \centering
    \includegraphics[width=2\columnwidth]{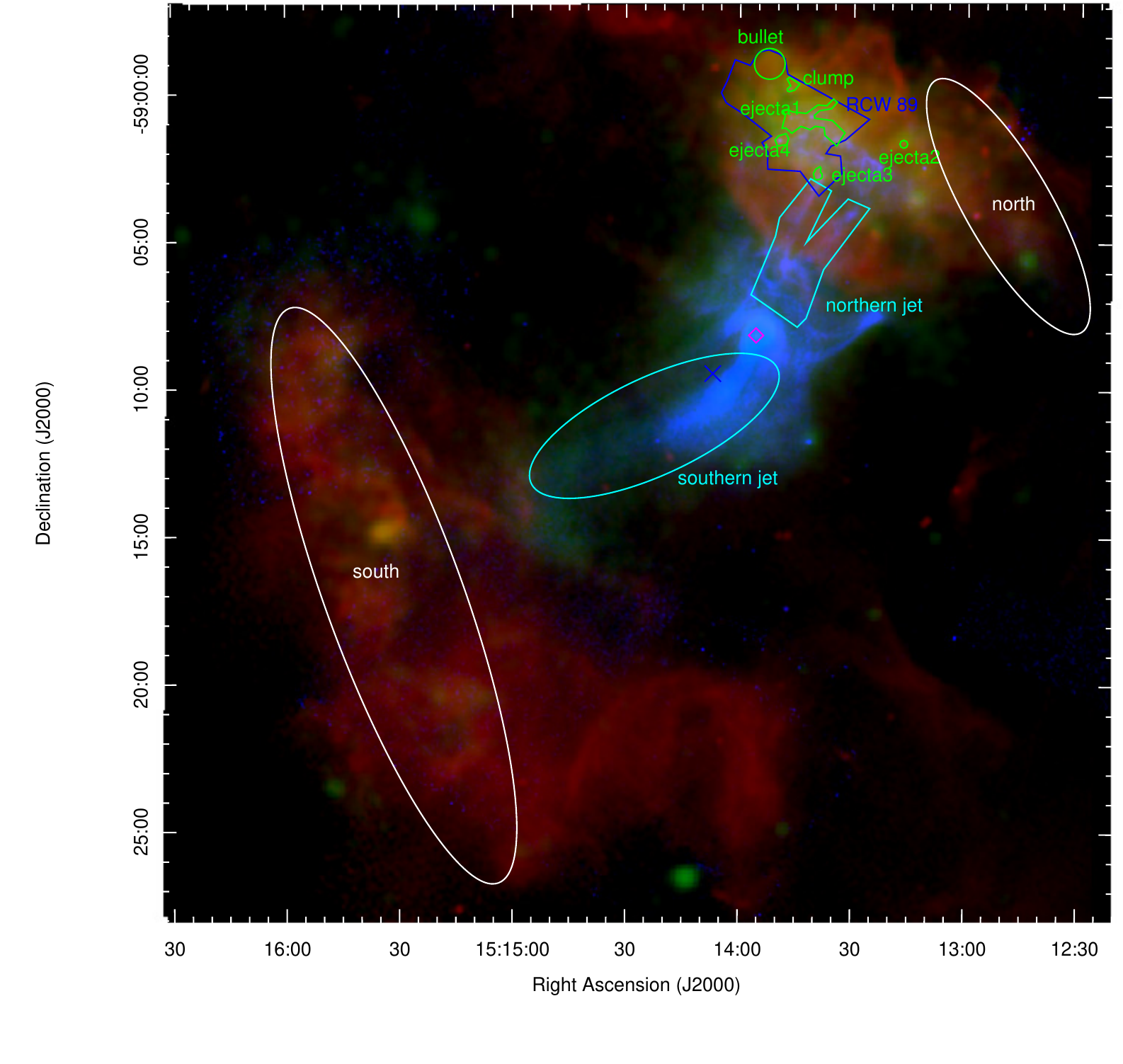}
    \caption{All regions used in this study overlaid on a multi‑wavelength image of MSH~15--52 (red: \textit{MeerKAT} image, green: 500--2000 eV \textit{eROSITA} image, blue: 3000--8000 eV \textit{Chandra} image). The blue cross indicates the centroid of the HESS $\gamma$-ray emission, which was derived by fitting the HESS data with a Gaussian profile \citep{Aharonian_2005}. The magenta diamond indicates the position of the pulsar, PSR~J1513--5908.}
    \label{fig:allregion}
\end{figure*}

We also investigated two regions shown in cyan in Figure \ref{fig:allregion}, in order to better understand the influence of the PWN jets on RCW 89 and on the large-scale SNR environment north and south of the PWN. The southern jet and potential northern jet regions (discussed below in Section~\ref{sec:jets}) have been observed many times with \textit{Chandra}, and our analysis utilized nine different observations taken between 2004 and 2022 with ObsIDs 3833, 4384, 5534, 5535, 6116, 6117, 20910, 20932, and 27448. 

We created a merged image of all the listed observations using \textit{merge\_obs}. Then, we used the ``Contour binning and accumulative smoothing software", also called ``contbin" \citep{Contbin}, to smooth the merged image and create binmaps based on variations in surface brightness. Next, we applied the regions defined by the binmaps to each observation and extracted spectra for the corresponding spatial bins. Spectra from the individual observations were subsequently fitted simultaneously, allowing us to construct the first photon-index maps of the full jet regions using multiple observations.

\begin{figure*}[htbp]
    \centering
    \includegraphics[width=2\columnwidth]{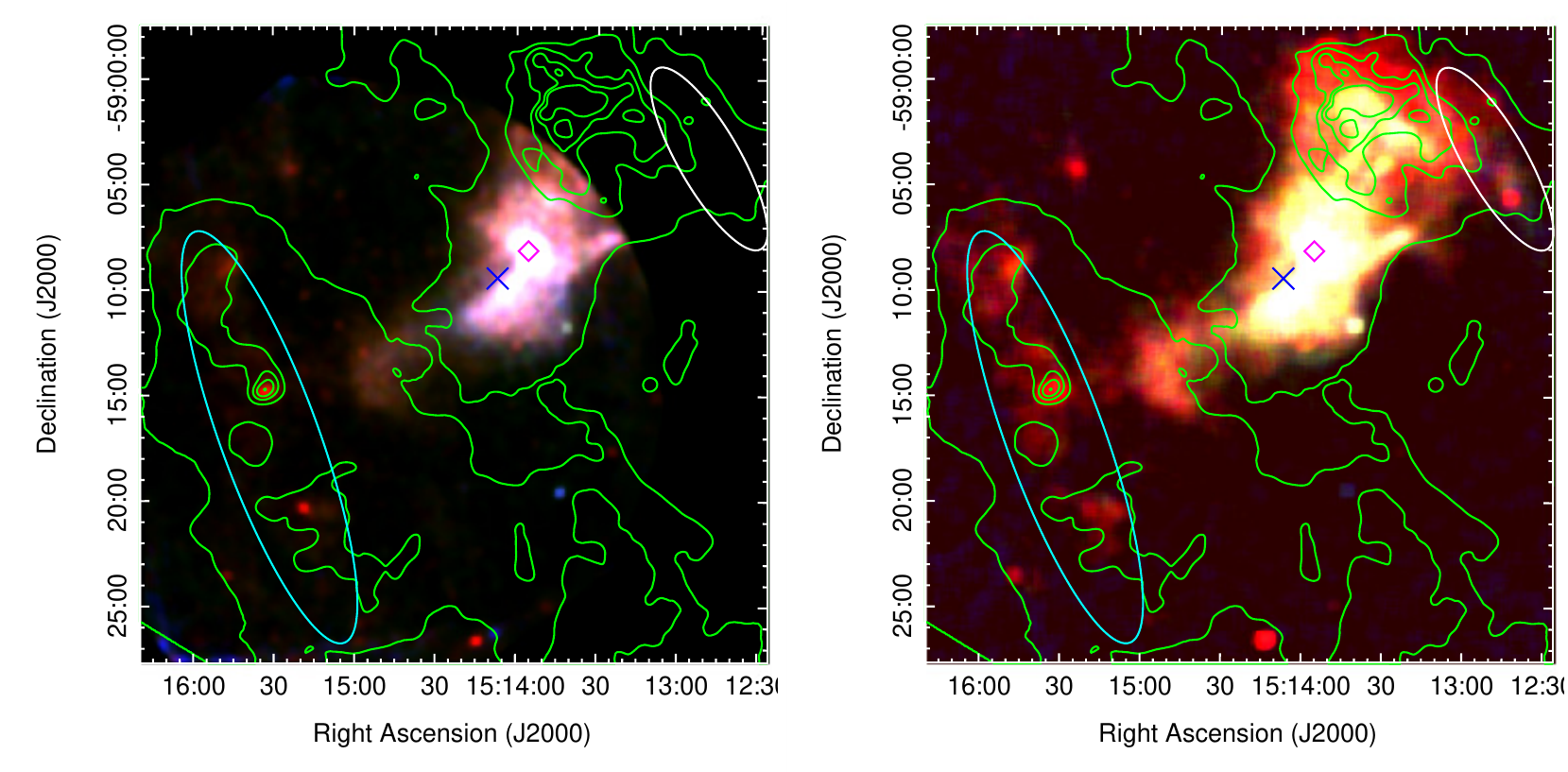}
    \caption{
    Tri-color \textit{XMM-Newton} (left) and \textit{eROSITA} (right) images of the southeastern region of MSH~15--52 with radio contours from \textit{MeerKAT} (red (500--2000 eV), green (2000--3000 eV), and blue (3000--8000 eV)). The cyan region corresponds to the proposed southern shell counterpart of MSH~15--52, while the white region marks the northern shell component. The blue cross indicates the centroid of the HESS $\gamma$-ray emission, which was derived by fitting the HESS data with a Gaussian profile \citep{Aharonian_2005}. The magenta diamond indicates the position of the pulsar, PSR~J1513--5908.
    }
    \label{fig:radioeroxmm}
\end{figure*}

\subsection{\textit{XMM-Newton}}

\textit{XMM-Newton} has observed MSH~15--52 and its associated PWN. However, due to strong flares, the photon counts were too low to analyze the properties of the clumps. Nevertheless, the larger field of view of \textit{XMM-Newton} compared to \textit{Chandra} makes it well suited for probing the region south of
the PWN and identifying the missing southern counterpart to the SNR shell in X-rays (see left panel of Figure \ref{fig:radioeroxmm}).

\textit{XMM-Newton} observed the southeast region of MSH~15--52 twice, in 2004 and 2005, with observation IDs 0302730101 and 0207050101, respectively. Because the latter was affected by strong flare contamination, we could not use it for spectral analysis. We used both observations to construct the image, but only the former for the spectral analysis. The data were processed using HEASoft 6.34 and Science Analysis System (SAS) package 20.0.0 \citep{2014SAS}. The \textit{XMM-Newton} imaging and spectroscopic analysis is discussed jointly with \textit{eROSITA} (see Section~\ref{sec:shell}).

\subsection{\textit{eROSITA}}

\textit{eROSITA} has observed MSH~15--52 in the Early Data Release with observation IDs 900035 and 900043. Compared to \textit{Chandra} and \textit{XMM-Newton}, \textit{eROSITA} provides a much larger field of view, which allows us to better investigate the overall morphology of the SNR as well as its surrounding environment. In the Early Data Release, the relatively long exposure time covers the entire remnant, enabling a comprehensive analysis of its global structure (see right panel of Figure \ref{fig:radioeroxmm}). In contrast, the \textit{eROSITA}-DE Data Release 1 offers an even wider coverage area, which is particularly useful for studying the environment around MSH~15--52. We used the observation IDs of 150226 and 150231.

We processed the data using the \textit{eROSITA} Science Analysis Software System (eSASS). Specifically, we applied the command \textit{evtool} to generate images, \textit{expmap} to create the exposure maps, and then calculated the flux maps. The spectra were extracted using \textit{srctool}. 
{Unless otherwise stated, all X-ray spectra presented in this work were fitted over the 0.4--7.0 keV energy range using XSPEC 12.14.1.}

In addition to the X-ray observations listed in Table~\ref{tab:xraylog}, we used radio data obtained with \textit{MeerKAT} and \textit{ASKAP} to provide a complementary multi-wavelength view of the SNR shell and its environment.

\section{results}
\label{sec:result}

\subsection{The ejecta region}
\label{sec:ejecta}

In the region of RCW 89, there have been many observations using \textit{Chandra}. In \cite{Borkowski_2020}, it was proposed that RCW 89 contains numerous ejecta clumps moving at high velocities. We analyzed the spectrum of the thermal emission region of RCW 89, where the hot plasma is produced by shock heating. To model the thermal emission, we considered three \textit{XSPEC} models:
\textit{apec}, a collisional ionization equilibrium (CIE) model for diffuse gas characterized by a single electron temperature; \textit{vpshock}, a non-equilibrium ionization (NEI) model for a plane-parallel shocked plasma with a constant temperature and a range of ionization timescales; and  \textit{vnei}, an NEI collisional plasma model with a single electron temperature and a single ionization parameter.
{We initially performed the spectral fitting with all elemental abundances fixed at their solar values. We then allowed the abundances of individual elements to vary to assess whether they could be meaningfully constrained by the data and whether their inclusion improved the fit. For Ni, its abundance was tied to that of Fe. No prominent emission lines associated with S, Ar, or Ca are apparent in the spectra, and allowing their abundances to vary independently does not result in a significant improvement in the fit. We therefore fixed the abundances of S, Ar, and Ca, as well as those elements whose emission lines are too weak to provide meaningful constraints, to 1.0 (solar).
}

\begin{figure*}[htbp]
	\centering
	\subfloat[RCW 89]{\includegraphics[height=\columnwidth,angle=-90]{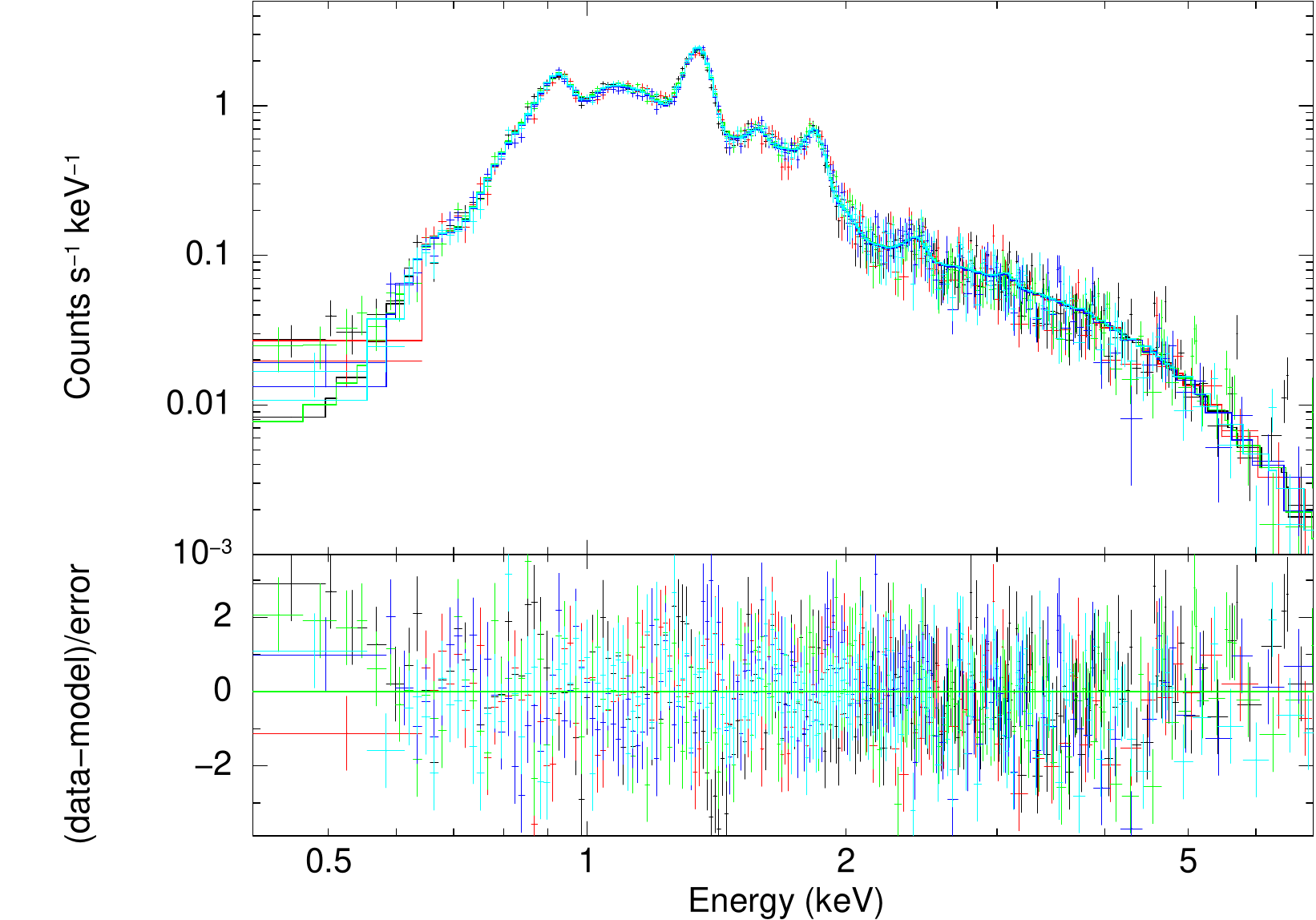}
		\label{fig:RCW89spec}}
	\subfloat[ejecta]{\includegraphics[height=\columnwidth,angle=-90]{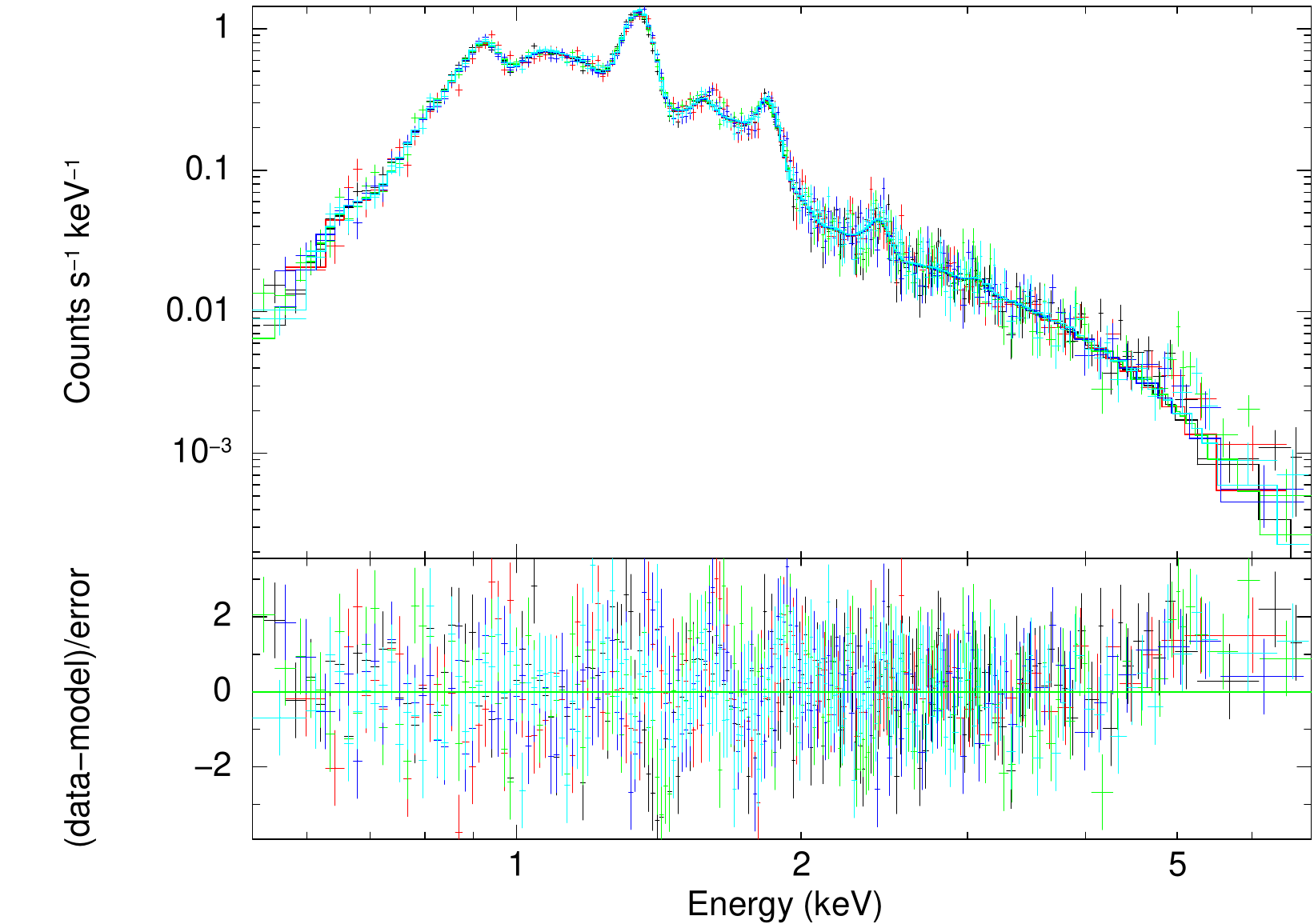}
		\label{fig:ejectaspec}}
	\caption{\textit{Chandra} spectra of the RCW 89 region (left) and the ejecta region (right).
    Black, red, green, blue, and cyan correspond to ObsIDs 19299, 19300, 20910, 20932, and 20933, respectively.
    All spectra are fitted simultaneously with the model \textit{TBabs}(\textit{vashift} $\times$ \textit{vpshock} + \textit{apec}).
    The lower panels show the residuals of the spectral fit normalized by their statistical errors ($delchi$ in \textit{XSPEC}).
    The solid line shows the model fitted to the data.}
	\label{fig:chandraspec1}
\end{figure*}

Motivated by previous reports of high-velocity ejecta in RCW~89, we adopted an absorbed \textit{vpshock} model convolved with \textit{vashift} to account for possible Doppler shifts in the emitting plasma. In the \textit{vashift} model, positive velocities correspond to redshifts, i.e., motion away from the observer, whereas negative velocities correspond to blueshifts, i.e., motion toward the observer. A single-component fit did not adequately reproduce the spectrum. Adding a non-thermal \textit{powerlaw} component also failed to improve the fit significantly. Finally, introducing an additional low-temperature thermal component yielded a better fit, as shown in Figure~\ref{fig:RCW89spec}.

\begin{figure*}[htbp!]
    \centering
    \includegraphics[width=1.95\columnwidth]{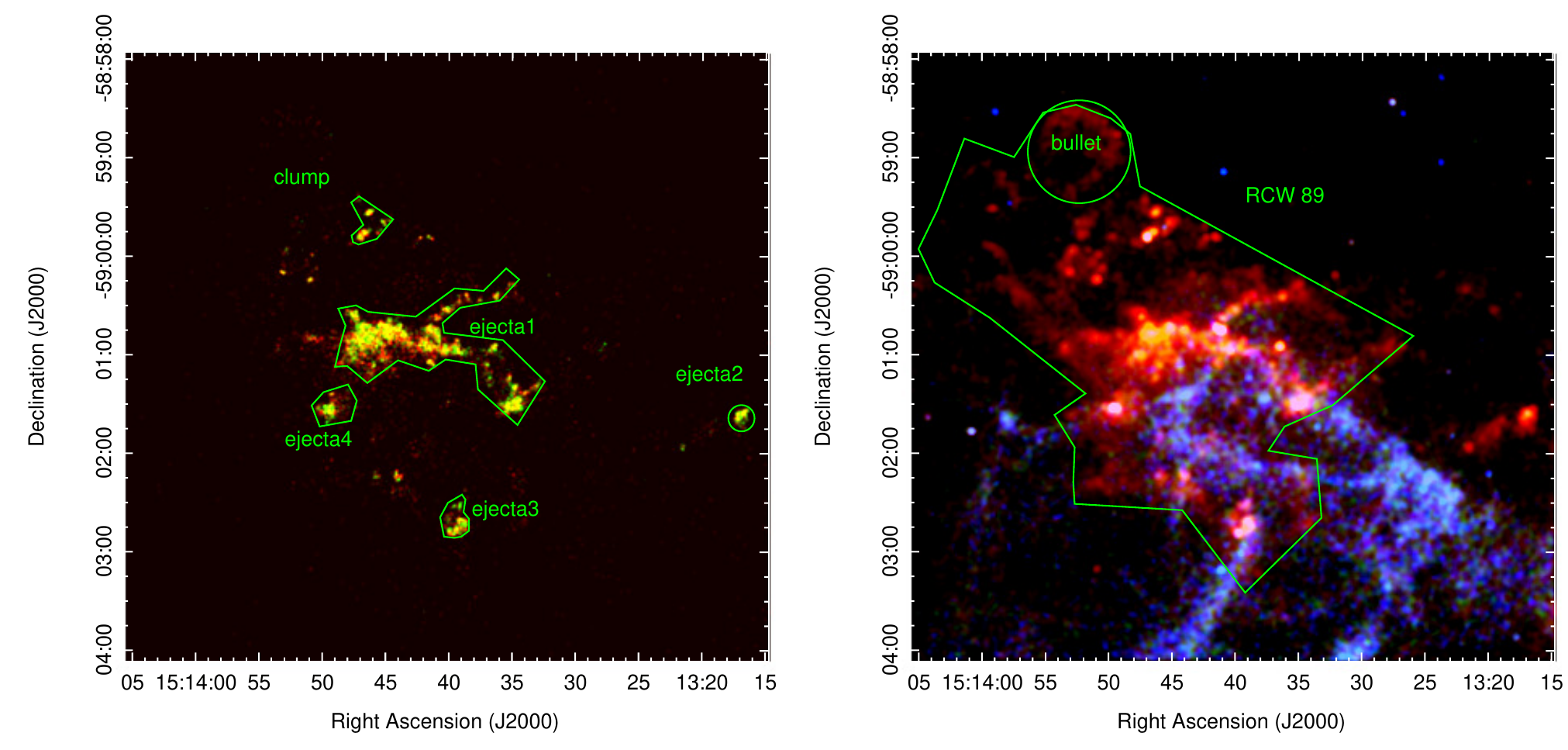}
    \caption{
   \redbf{(Left): Equivalent Width (EW) map of RCW~89 highlighting the emission from Ne (red) and Mg (green). (Right): Tri‑color X‑ray image of RCW~89, with red, green, and blue corresponding to the 500--2000 eV, 2000--3000 eV, and 3000--9000 eV energy bands, respectively. 
   Both images are derived from the \textit{Chandra/ACIS} data. 
   The regions overlaid are those selected for our spatially resolved spectroscopic study (see Section~\ref{sec:ejecta}).
    }}
    \label{fig:ewimage}
\end{figure*}

The overall spectrum shows strong emission lines for Ne and Mg. 
{Therefore, we constructed equivalent width (EW) maps for these two lines separately to trace potential ejecta regions, as shown in the left panel of Figure~\ref{fig:ewimage}. For each line, we selected a narrow target band (Mg: 1.25--1.5 keV; Ne: 0.86--0.98 keV) and two adjacent continuum bands (Mg: 1.1--1.2 and 1.5--1.7 keV; Ne: 0.82--0.86 and 0.98--1.1 keV). For each line, we generated flux images for the target and continuum bands, subtracted the averaged continuum from the target to obtain the net line flux, and then divided by the continuum and multiplied by the target band's energy width to produce the EW map pixel by pixel. Negative values were set to zero.}

{The EW map regions correspond to clumpy structures in the soft X-ray band. We extracted spectra from several distinct regions, including ``ejecta1'', ``ejecta2'', ``ejecta3'', ``ejecta4'', a shocked‑ISM clump, and a bullet‑like structure (see Figure~\ref{fig:ewimage}, Table~\ref{tab:clump} and Table~\ref{tab:ejecta}). We note that for certain regions and elements, the abundances could not be freely constrained by the data; in those cases, we adopted fixed values with clear justifications. Specifically, for the bullet‑like region, the O abundance could not be reliably determined, freeing it returned unphysical values, while fixing it to solar produced large residuals, so we fixed O to the value derived from the ``RCW~89'' spectrum; the Si abundance in this region was also poorly constrained and thus fixed to 1.0 (solar). For ``ejecta1'', the strong low‑energy absorption and the blending of Fe and O lines prevented a meaningful Fe abundance measurement, so we fixed Fe to the global ``RCW~89'' value as well. 
In contrast, for ``ejecta2'', ``ejecta3'', and ``ejecta4'', the absorption column density ${N}_{\mathrm{H}}$ was already fixed, which allowed Fe and O to be left free, yielding the uncertainties reported in Table~\ref{tab:ejecta}. 
The spectrum of region ``ejecta1'' is shown in Figure~\ref{fig:ejectaspec}.
With these constraints, we find that ``ejecta2'', ``ejecta3'', and ``ejecta4'' exhibit high velocities and enhanced abundances, and therefore identify them as ejecta clumps. We assume these clumps consist of pure ejecta and compare them with core‑collapse nucleosynthesis models in Section~\ref{subsec:progenitor}. In contrast, the ``clump'' shows lower velocity and reduced abundance, which we interpret as shocked ISM. 
\redbf{
Given its high Ne abundance and distinct morphology, we interpret the bullet-like region}
as a fast-moving ejecta structure. The spectra for the regions ``ejecta2'', ``ejecta3'', ``ejecta4'', and ``clump'' are presented in Figure~\ref{fig:chandraspec2} in the appendix.}

We note that the velocities reported in Tables~\ref{tab:clump} and \ref{tab:ejecta} are line-of-sight velocities obtained from the \textit{vashift} component, rather than total space velocities. In the \textit{vashift} model, the model energies are shifted by \(\Delta E=-Ev/c\), such that positive velocities correspond to redshifts (motion away from the observer), while negative velocities correspond to blueshifts (motion toward the observer). For ejecta structures with corresponding proper motion measurements, the line-of-sight velocity can be combined with the radial and tangential velocities in the plane of the sky to estimate the three-dimensional space velocity as
\begin{equation}
v_{\rm 3D}=\sqrt{v_{\rm LOS}^{2}+v_r^{2}+v_t^{2}},
\end{equation}
where \(v_r\) and \(v_t\) are the radial and tangential velocities with respect to the pulsar reported by \cite{Borkowski_2020}. We find that ``ejecta2'', ``ejecta3'', and ``ejecta4'' spatially overlap with the Borkowski et al. regions G, F, and B, respectively. Under the assumption that the spectroscopic and proper motion measurements trace the same dominant structures, ``ejecta2'' and ``ejecta4'' have estimated three-dimensional velocities of \(5600\pm700\) km s\(^{-1}\) and \(4390\pm800\) km s\(^{-1}\), respectively. For ``ejecta2'', the positive \(v_{\rm LOS}\) indicates motion away from the observer, while the positive \(v_r\) indicates motion radially away from the pulsar in the plane of the sky. For ``ejecta4'', the corresponding positive \(v_{\rm LOS}\) likewise indicates motion away from the observer. The resulting velocity vectors are inclined by approximately \(23^\circ\) and \(28^\circ\), respectively, from the plane of the sky toward the far side of the remnant. A unique three-dimensional velocity cannot be assigned to ``ejecta3'', because the corresponding \cite{Borkowski_2020} region F contains the distinct knots F$_{\rm s}$ and F$_{\rm f}$, whose radial proper motion velocities differ substantially (\(760\pm140\) and \(4190\pm170\) km s\(^{-1}\), respectively).
Likewise, ``ejecta1'' is a substantially larger extraction region without a one-to-one correspondence to an individual proper motion knot. We therefore retain the line-of-sight velocities from our spectral fits for these regions rather than assigning uncertain three-dimensional velocities.

We also compare our spectral parameters with those reported by \cite{Borkowski_2020}. Their analysis of bright fast-moving ejecta knots yielded \(N_{\rm H}=1.32^{+0.11}_{-0.10}\times10^{22}\)~cm\(^{-2}\), \(kT=0.57^{+0.44}_{-0.20}\)~keV, and an ionization timescale of \(1.03^{+0.84}_{-0.65}\times10^{11}\)~cm\(^{-3}\)~s, together with a blueshifted velocity of \(3100^{+700}_{-600}\)~km~s\(^{-1}\) (90\% confidence intervals). Their \(N_{\rm H}\) is consistent with the value adopted in our ejecta1--ejecta4 fits, while their best-fit temperature is lower than those obtained for our ejecta regions, although their 90\% confidence interval overlaps with that of ``ejecta2'' (\(kT=1.2^{+0.6}_{-0.3}\)~keV). The relative abundance pattern of Ne, Mg, and Si is broadly consistent with our results, with enhanced Ne and Mg relative to Si. We do not directly compare the O abundance because it varies substantially among our extraction regions and is strongly affected by blending with Fe, resulting in relatively large uncertainties. Moreover and importantly, the regions analyzed by \cite{Borkowski_2020} do not correspond one-to-one to any of our extraction regions. These differences in extraction regions and spectral models preclude a direct one-to-one comparison of individual spectral parameters, but the overall consistency in absorption, temperature, velocity scale, and abundance pattern supports our identification of the high-velocity regions as ejecta clumps.

\begin{table*}[htbp]
\centering
\caption{{Best-fit results and 90\% uncertainties for the regions ``RCW~89'', ``clump'', and ``bullet'' (see Figure~\ref{fig:ewimage}). 
The RCW~89 spectrum is fitted with a two-component absorbed model consisting of a \textit{vpshock} component plus an additional \textit{apec} soft component, with the absorption column density left free. 
The ``clump'' and ``bullet'' spectra are fitted with a single \textit{vpshock} component, and their absorption column densities are fixed to the value derived for the ``RCW~89'' region.}
}
\label{tab:clump}
\begin{tabular}{lccc}
\hline\hline
 & RCW 89 & clump & bullet \\
\hline\hline
${N}_{\mathrm{H}}\ (\rm{10^{22}\ cm^{-2}}$) & $1.30^{+0.05}_{-0.05}$ & 1.30 {(fixed)} & 1.30 {(fixed)} \\
{$\mathit{velocity \ \mathrm{(km/s)}}^{\dagger}$} 
& $1500^{+200}_{-200}$ 
& $-300^{+500}_{-400}$ 
& $4000^{+2000}_{-2000}$ \\
$\textit{kT\ \rm{(keV)}}$ & $2.38^{+0.12}_{-0.11}$ & $0.43^{+0.12}_{-0.06}$ & $0.34^{+0.11}_{-0.10}$ \\
\textit{O} & $0.7^{+0.8}_{-0.6}$ & $0.3^{+0.4}_{-0.3}$ & 0.69 {(fixed)} \\
\textit{Ne} & $2.0^{+0.4}_{-0.3}$ & $1.1^{+1.0}_{-0.3}$ & $1.5^{+0.9}_{-0.8}$ \\
\textit{Mg} & $2.1^{+0.1}_{-0.1}$ & $1.4^{+1.0}_{-0.3}$ & $1.0^{+0.5}_{-0.4}$ \\
\textit{Si} & $0.66^{+0.05}_{-0.05}$ & $0.6^{+0.2}_{-0.1}$ & 1.00 {(fixed)} \\
\textit{Fe} & $0.6^{+0.2}_{-0.3}$ & $0.12^{+0.10}_{-0.08}$ & $0.7^{+0.4}_{-0.4}$ \\
$\tau$~($10^{10}$  $\rm cm^{-3}\ s)$ & $2.2^{+0.1}_{-0.1}$ & $27^{+27}_{-13}$ & {$<270$} \\
$\textit{norm}\ (10^{-3} \ \rm{cm^{-5}})$ & $9.0^{+0.5}_{-0.5}$ & $4.1^{+3.1}_{-4.1}$ & $3.5^{+10.1}_{-2.1}$ \\
$\textit{apec kT\ \rm{(keV)}}$ & $0.19^{+0.01}_{-0.01}$ & - & - \\
$\textit{apec norm}\ (\rm{cm^{-5}})$ & $0.39^{+0.10}_{-0.08}$ & - & - \\
$\chi^2_{\nu}$(d.o.f.) & 1.189(1867) & 1.150(355) & 0.906(117) \\
\hline
\end{tabular}

\textit{Note.} $^{\dagger}$ Velocities are line-of-sight velocities derived from the \textit{vashift} component. 
Positive values correspond to redshifted (receding) plasma, while negative values correspond to blueshifted (approaching) plasma. 
This sign convention is used for all velocity measurements reported in this work.

\end{table*}

\begin{table*}[htbp]
\centering
\caption{{Best-fit results and 90\% uncertainties for the ejecta regions (see Figure~\ref{fig:ewimage}). 
The ``ejecta1'' spectrum is fitted with a two-component absorbed model consisting of a \textit{vpshock} component plus an additional \textit{apec} soft component, with the absorption column density left free. 
The spectra of ``ejecta2'', ``ejecta3'', and ``ejecta4'' are fitted with a single \textit{vpshock} component, and their absorption column densities are fixed to the value derived for the ``RCW~89'' region.} 
}
\label{tab:ejecta}
\begin{tabular}{lcccc}
\hline\hline
 & ejecta1 & ejecta2 & ejecta3 & ejecta4 \\
\hline\hline
${N}_{\mathrm{H}}\ (\rm{10^{22}\ cm^{-2}}$) & $1.34^{+0.04}_{-0.04}$ & 1.30 {(fixed)} & 1.30 {(fixed)} & 1.30 {(fixed)} \\
{$\mathit{velocity \ \mathrm{(km/s)}}$} & $1400^{+200}_{-200}$ & $2200^{+700}_{-600}$ & $2800^{+600}_{-800}$ & $2000^{+800}_{-800}$ \\
$\textit{kT\ \rm{(keV)}}$ & $1.46^{+0.07}_{-0.06}$ & $1.2^{+0.6}_{-0.3}$ & $2.0^{+0.4}_{-0.3}$ & $1.5^{+0.2}_{-0.2}$ \\
\textit{O} & $1.5^{+1.0}_{-0.8}$ & $0.8^{+8.3}_{-0.8}$ & $10^{+4}_{-3}$ & $12^{+5}_{-3}$ \\
\textit{Ne} & $3.6^{+0.5}_{-0.5}$ & $7^{+10}_{-3}$ & $5^{+2}_{-2}$ & $6.2^{+2.2}_{-1.6}$ \\
\textit{Mg} & $3.7^{+0.3}_{-0.3}$ & $10^{+10}_{-4}$ & $5.9^{+1.5}_{-1.0}$ & $5.9^{+1.7}_{-1.2}$ \\
\textit{Si} & $1.2^{+0.1}_{-0.1}$ & $1.8^{+1.9}_{-0.8}$ & $1.3^{+0.4}_{-0.3}$ & $1.6^{+0.5}_{-0.4}$ \\
\textit{Fe} & 0.56 {(fixed)} & $2.3^{+2.1}_{-1.6}$ & $4.0^{+1.9}_{-1.5}$ & $1.7^{+1.4}_{-1.1}$ \\
$\tau$~($10^{10}$  $\rm cm^{-3}\ s)$ & $2.8^{+0.2}_{-0.2}$ & $2.7^{+0.8}_{-0.7}$ & $3.4^{+0.6}_{-0.5}$ & $3.8^{+0.7}_{-0.5}$ \\
$\textit{norm}\ (10^{-3} \ \rm{cm^{-5}})$ & $3.6^{+0.3}_{-0.3}$ & $0.09^{+0.11}_{-0.05}$ & $0.15^{+0.04}_{-0.03}$ & $0.22^{+0.07}_{-0.06}$ \\
$\textit{apec kT\ \rm{(keV)}}$ & $0.20^{+0.01}_{-0.01}$ & - & - & - \\
$\textit{apec norm}\ ( \rm{cm^{-5}})$ & $0.13^{+0.04}_{-0.03}$ & - & - & - \\
$\chi^2_{\nu}$(d.o.f.) & 1.300(1022) & 0.951(170) & 1.001(351) & 1.104(397) \\
\hline
\end{tabular}

\end{table*}

\begin{figure}[htbp]
    \centering
    \includegraphics[height=\columnwidth,angle=-90]{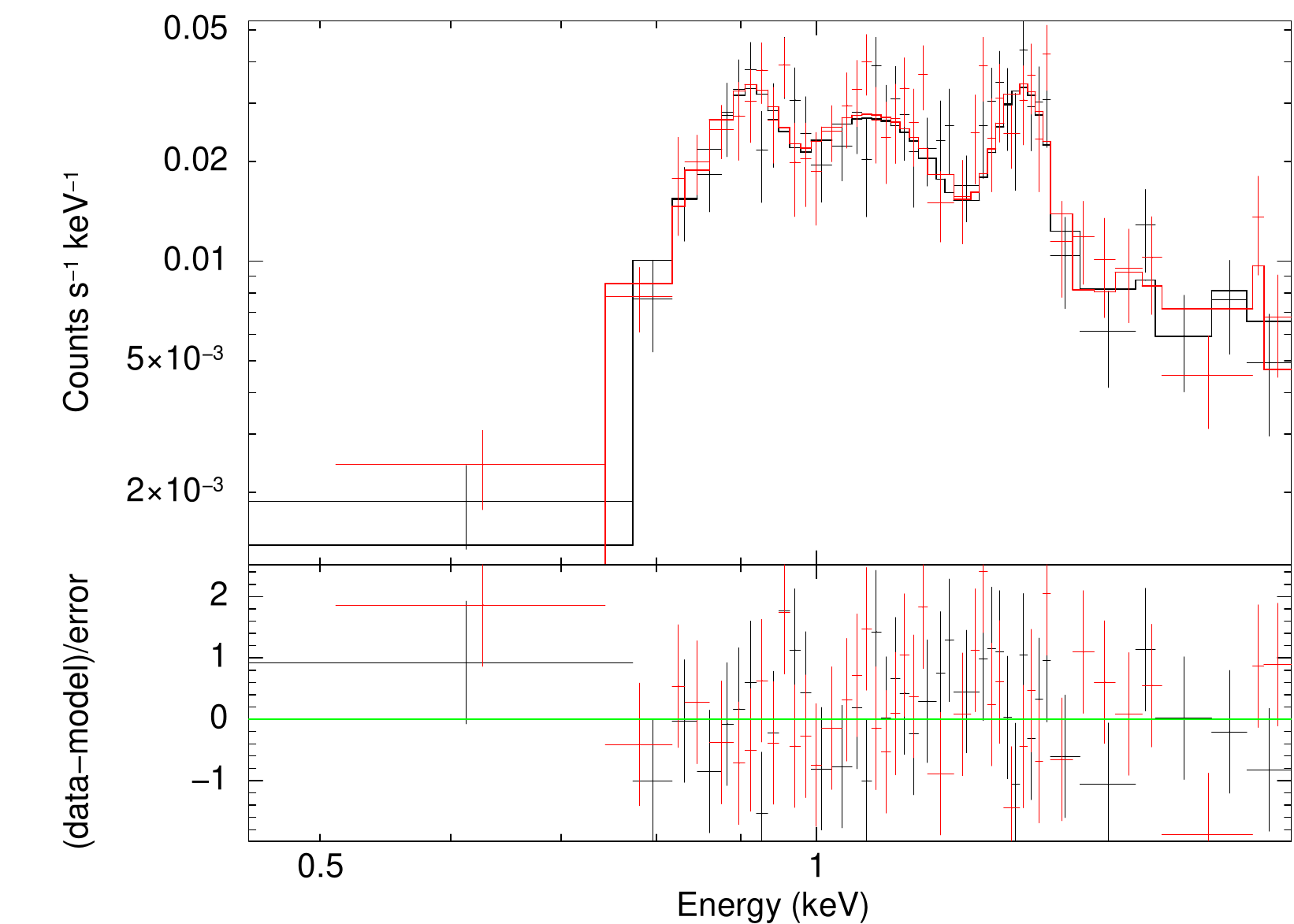}
    \caption{\textit{Chandra} spectrum of the ``bullet'' region. Black and red correspond to ObsIDs 19299 and 20910, respectively. Spectrum is fitted with the model \textit{TBabs}(\textit{vashift}$\times$\textit{vpshock}).}
    \label{fig:chandrabulletspec}
\end{figure}

\subsection{The jet regions}
\label{sec:jets}

The southern and northern jets have been previously analyzed, including early \textit{Chandra} data \citep{DeLaney_2006}, \textit{IXPE} data \citep{Romani_2023}, and \textit{NuSTAR} with \textit{Chandra} data \citep{An_2014}. These studies analyzed different jet regions and reported results of their absorbed power-law fits. In \cite{DeLaney_2006} and \cite{Romani_2023}, the column densities for each fit, $N_{H}$, are frozen to similar values of 0.86 and 0.896 $\times\ 10^{22}$ cm$^{-2}$, respectively. \cite{An_2014}, however, reported fits with varying column density.

\begin{figure}[ht!]
    \centering    \includegraphics[width=\columnwidth]{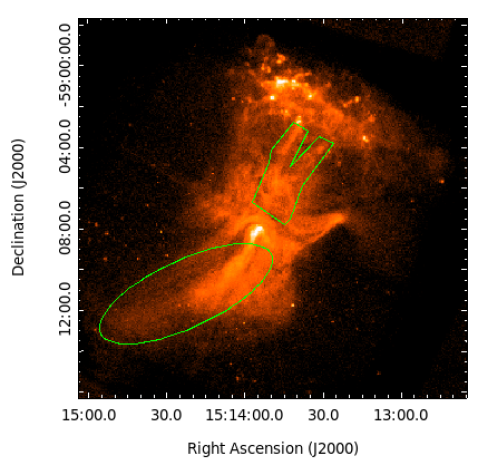}
    \caption{Merged \textit{Chandra} observations of the B1509--58 PWN and RCW 89. Overlay of the southern jet and northern jet regions, which were investigated to search for variations in the photon index as seen in Figure \ref{fig:photonIndex_Jet}.}
    \label{fig:Chandra_mapRegs}
\end{figure}

\begin{figure*}
	\centering
	\subfloat[{northern jet region}]{\includegraphics[width=\columnwidth]{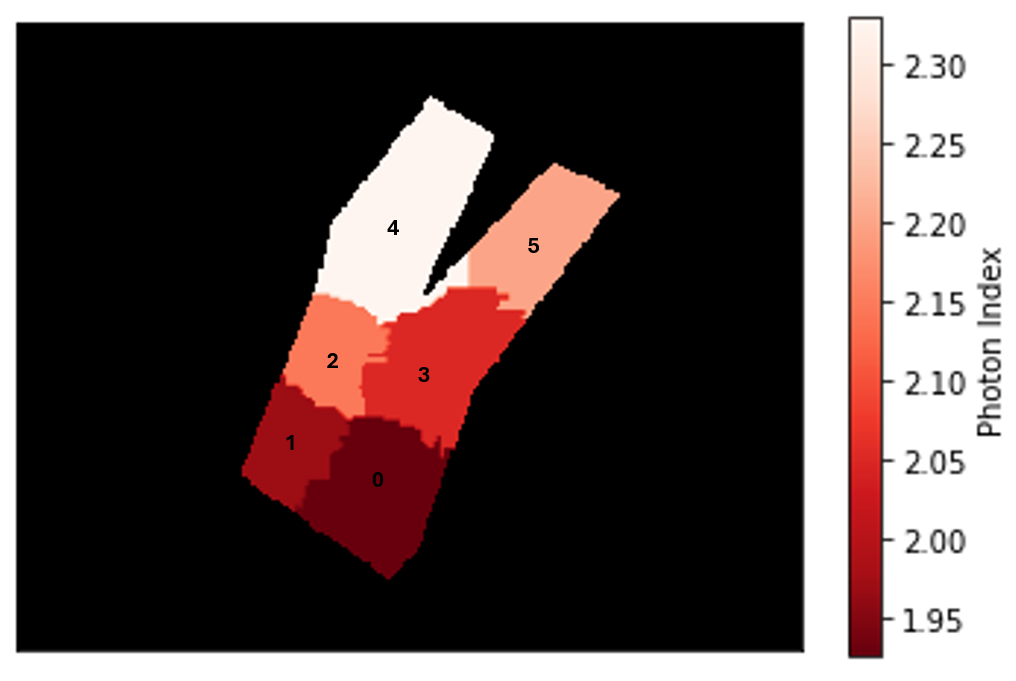}
		\label{fig:photonIndex_pants}}
	\subfloat[{southern jet region}]{\includegraphics[width=\columnwidth]{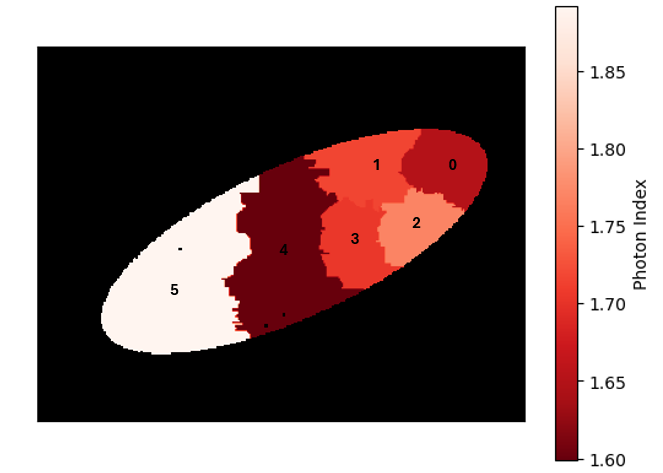}
		\label{fig:photonIndex_southJet}}
	\caption{
    \textit{Chandra} photon-index map of the potential northern jet region (left) and southern jet region (right). Regions are labeled 0 to 5 (see Figure~\ref{fig:allregion} and Figure~\ref{fig:Chandra_mapRegs} context). {The uncertainties are listed in Table \ref{tab:alljetregionfit}.
    }}
	\label{fig:photonIndex_Jet}
\end{figure*}

In our study of the northern and southern jet regions shown in Figure \ref{fig:Chandra_mapRegs}, we performed fits with both frozen and varying $N_{H}$, and each resulted in similar reduced chi-squared values. With the assumption that there is no foreground source of absorption, and the comprehensive results from \cite{An_2014} on varying $N_{H}$ in \textit{Chandra} fits, we focus on the frozen column density fit results in this paper. In our analysis, each region was fit with an absorbed power-law model, \textit{tbabs $\times$ powerlaw}, with column density frozen to values from the full-region simultaneous fit of every observation. The northern jet region is fitted with $N_{H}$ $= 1.20 \times 10^{22}$ cm$^{-2}$, and the southern jet region with $N_{H}$ $= 1.22\times 10^{22}$ cm$^{-2}$. These column density values are higher than previous analyses, possibly due to the use of a more updated absorption model.
The photon index maps resulting from fits with these column densities are shown in Figures \ref{fig:photonIndex_pants} and \ref{fig:photonIndex_southJet}. 

\begin{table*}[ht!]
\centering
\caption{Best-fit results and 90\% uncertainties for all jet regions. All $N_{H}$ values are frozen to 1.2$\times$10$^{22}$~cm$^{-2}$.}
\label{tab:alljetregionfit}
\begin{tabular}{llccc}
\hline\hline
Jet & Region & norm (10$^{-3}$ cm$^{-5}$) & Photon Index & $\chi^2_\nu$(d.o.f.)\\
\hline
South & Region 0 & $1.57^{+0.02}_{-0.02}$ & $1.65^{+0.01}_{-0.01}$ & 1.78(1180)\\
      & Region 1 & $1.01^{+0.02}_{-0.02}$ & $1.72^{+0.02}_{-0.02}$ & 1.17(975)\\
      & Region 2 & $1.16^{+0.02}_{-0.02}$ & $1.78^{+0.02}_{-0.02}$ & 1.27(983)\\
      & Region 3 & $1.05^{+0.02}_{-0.02}$ & $1.70^{+0.02}_{-0.02}$ & 1.36(983)\\
      & Region 4 & $0.98^{+0.02}_{-0.02}$ & $1.62^{+0.02}_{-0.02}$ & 1.06(1048)\\
      & Region 5 & $1.04^{+0.03}_{-0.03}$ & $1.91^{+0.03}_{-0.03}$ & 1.11(947)\\
\hline
North & Region 0 & $0.81^{+0.01}_{-0.01}$ & $1.92^{+0.02}_{-0.02}$ & 2.02(1344)\\
      & Region 1 & $0.60^{+0.03}_{-0.03}$ & $1.94^{+0.07}_{-0.07}$ & 1.01(975)\\
      & Region 2 & $0.73^{+0.01}_{-0.01}$ & $2.13^{+0.02}_{-0.02}$ & 1.22(1080)\\
      & Region 3 & $0.87^{+0.02}_{-0.02}$ & $2.04^{+0.02}_{-0.02}$ & 1.38(1294)\\
      & Region 4 & $1.34^{+0.02}_{-0.02}$ & $2.33^{+0.02}_{-0.02}$ & 1.20(1459)\\
      & Region 5 & $0.56^{+0.01}_{-0.01}$ & $2.18^{+0.03}_{-0.03}$ & 1.62(929)\\
\hline
\end{tabular}
\end{table*}

\subsection{The shell region}
\label{sec:shell}

The early \textit{eROSITA} observations cover the entire remnant and reveal clear shell structures in both the northwest and the southeast regions (see right panel of Figure \ref{fig:radioeroxmm}). We extracted spectra from both shells for detailed analysis. For the southern shell, the region is also covered by \textit{XMM-Newton} MOS2 and pn (see left panel of Figure \ref{fig:radioeroxmm}), allowing us to perform joint fits with \textit{eROSITA} data. In contrast, the northern shell is only covered by \textit{eROSITA}.
{A notable finding from the \redbf{\textit{eROSITA}} data is the clear detection of a shell-like thermal structure in the northern region. Its location lies significantly farther out than the blast-wave position inferred by \citet{Borkowski_2020} from Chandra proper‑motion measurements.
This implies that the full extent of the SNR shock may be larger than previously estimated. We discuss the implications of this discrepancy for the evolution and morphology of RCW 89 in Section~\ref{sec:properityofsnr}.}

\begin{figure*}
	\centering
	\subfloat[vpshock]{\includegraphics[height=\columnwidth,angle=-90]{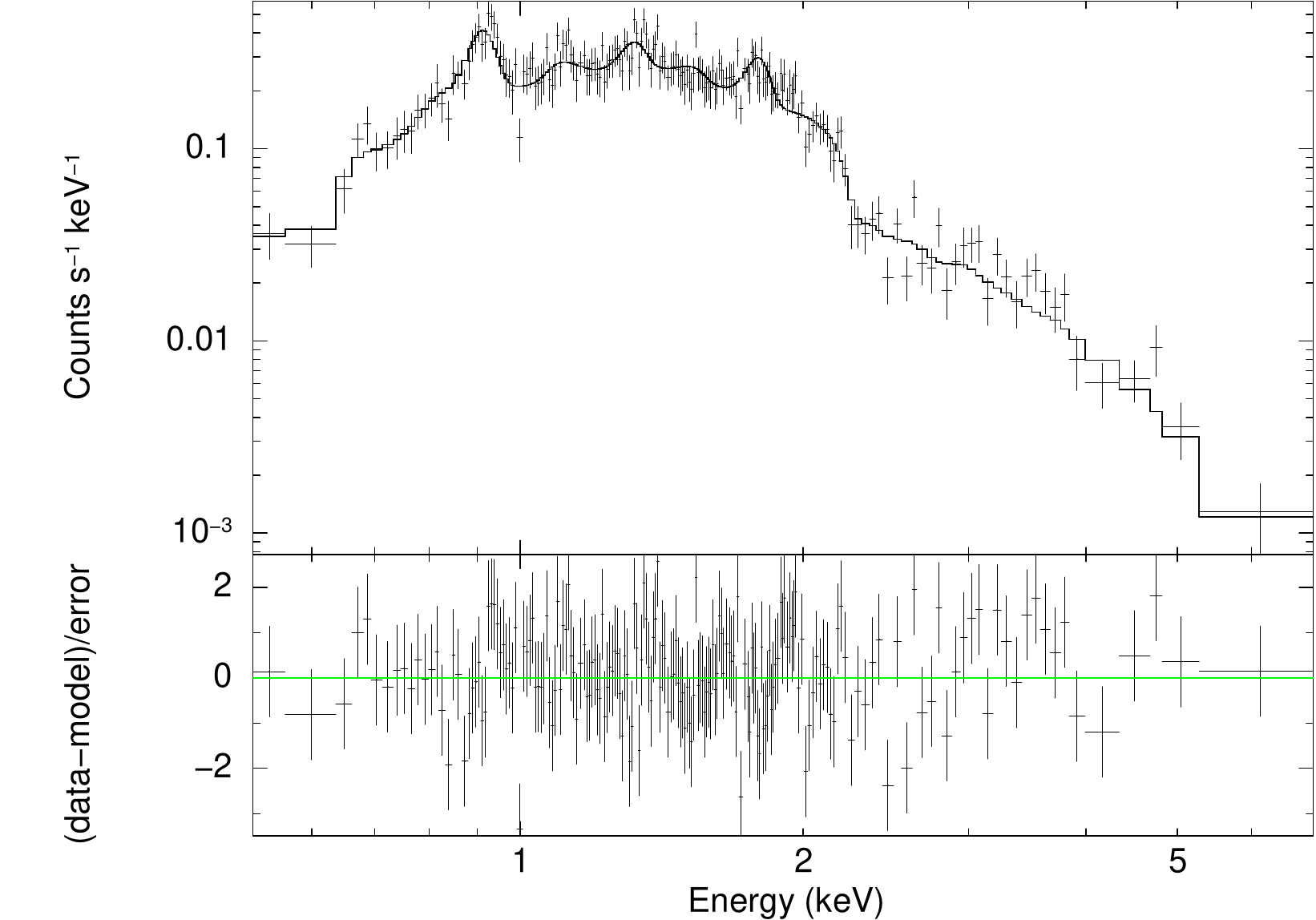}
		\label{fig:norththvpshock}}
	\subfloat[vpshock+powerlaw]{\includegraphics[height=\columnwidth,angle=-90]{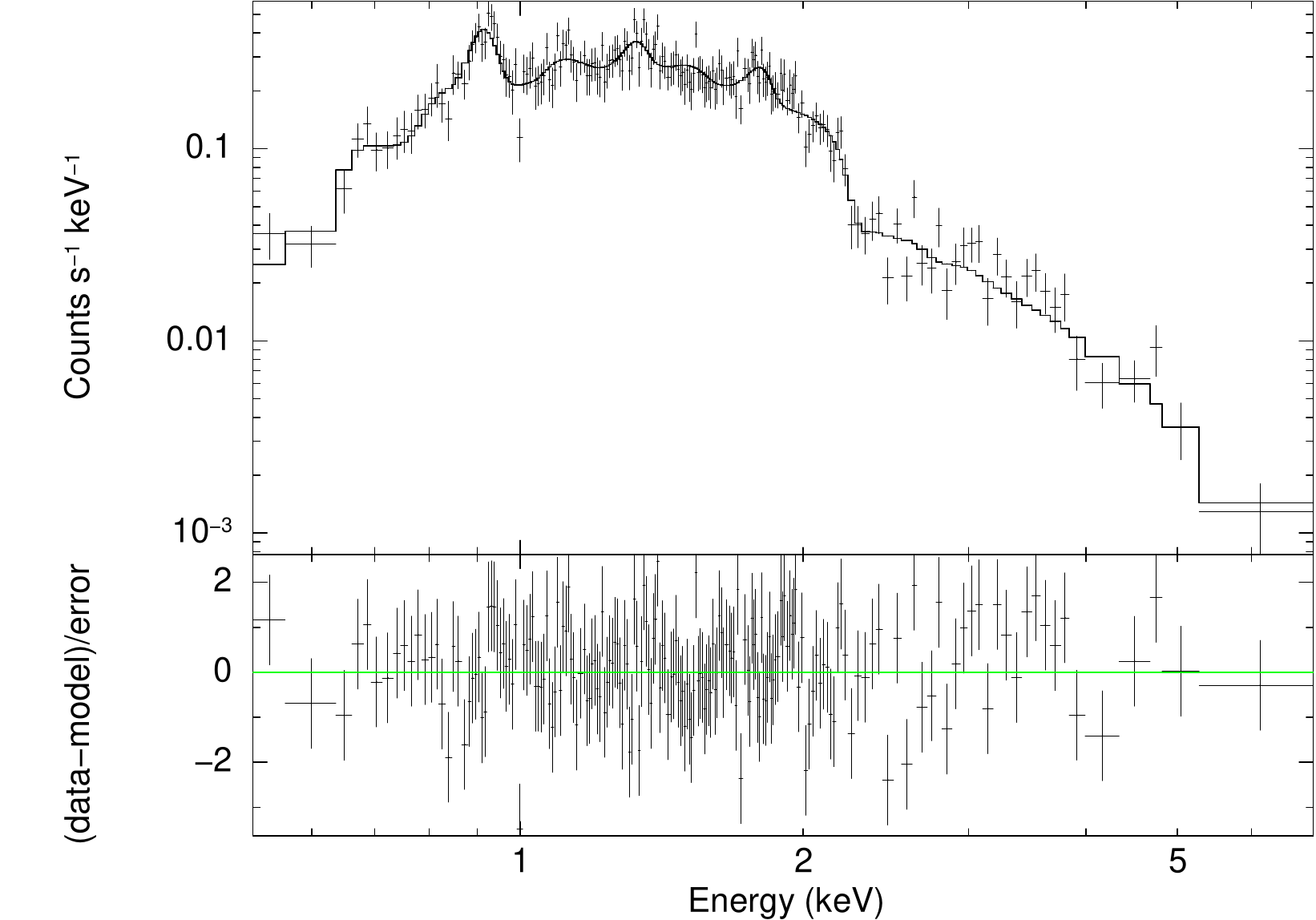}
		\label{fig:northvpshockpow}}
	\caption{\textit{eROSITA} spectrum of the northern shell with a single component \textit{vpshock} model (left) and with a two-component, \textit{vpshock+powerlaw}, model (right).}
	\label{fig:northshellspec}
\end{figure*}

For the northern shell, we considered thermal, non-thermal, and two-component hybrid models.
A pure power-law model failed to
adequately reproduce the northern shell spectrum. We find that the spectrum can instead be characterized either by a high-temperature, single-component \textit{vpshock} thermal model with $kT = 3.41$~keV, or by a two-component model consisting of a lower-temperature \textit{vpshock} component with $kT = 1.98$~keV plus a power-law component. The fitting results are summarized
in Table \ref{tab:shellspec} and the spectra for both scenarios are shown in Figure \ref{fig:northshellspec}.

For the southern shell, we performed joint fits using the \textit{eROSITA}, MOS2, and pn spectra.
As for the northern shell, a pure power-law model failed to reproduce the data, indicating that a thermal component is required. However, the resulting fits either remained statistically poor, 
with $\chi^2_{\nu} \sim 1.6$, or converged to unrealistically low absorption values ($N_{\mathrm{H}} \sim 0.7$ $\times$ 10$^{22}$~cm$^{-2}$), significantly below those inferred from the \textit{Chandra} observations.
These results suggest that the shell's spectrum could be affected by contamination from an additional foreground source along the line of sight.

\begin{figure}[htbp!]
    \centering
    \includegraphics[width=\columnwidth]{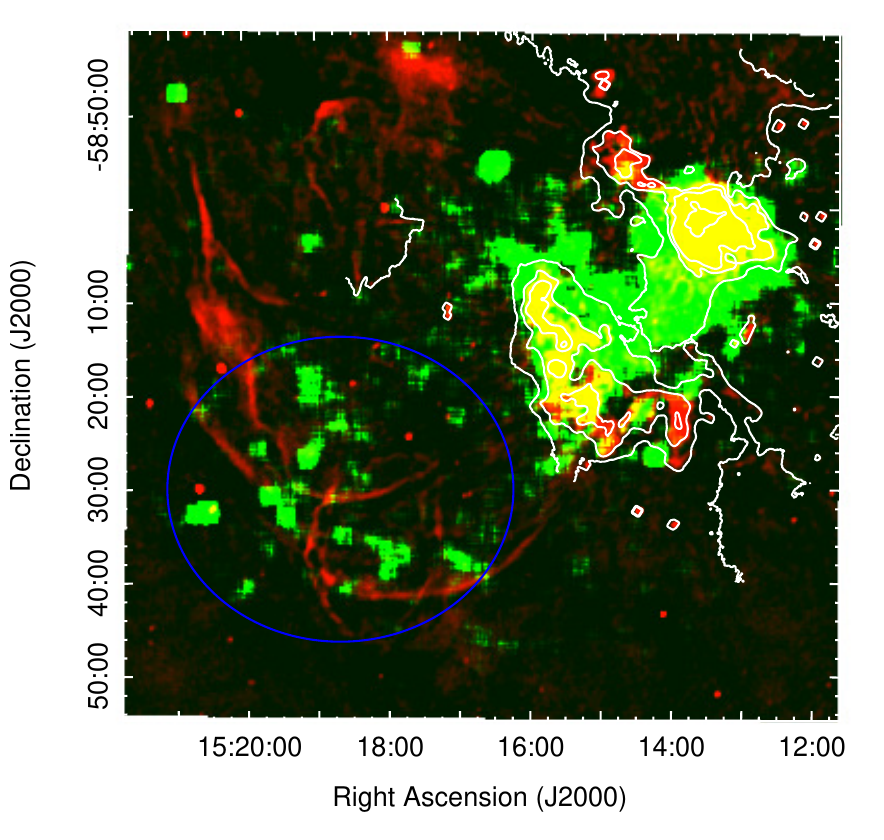}
    \caption{Multi-wavelength image of the region surrounding MSH~15--52.   Red: \textit{ASKAP} image, green: \textit{eROSITA} image, and  white contours showing the \textit{MeerKAT} data. The blue elliptical region highlights the foreground source.}
    \label{fig:erositadr1}
\end{figure}

\begin{figure}[htbp]
    \centering
    \includegraphics[height=\columnwidth,angle=-90]{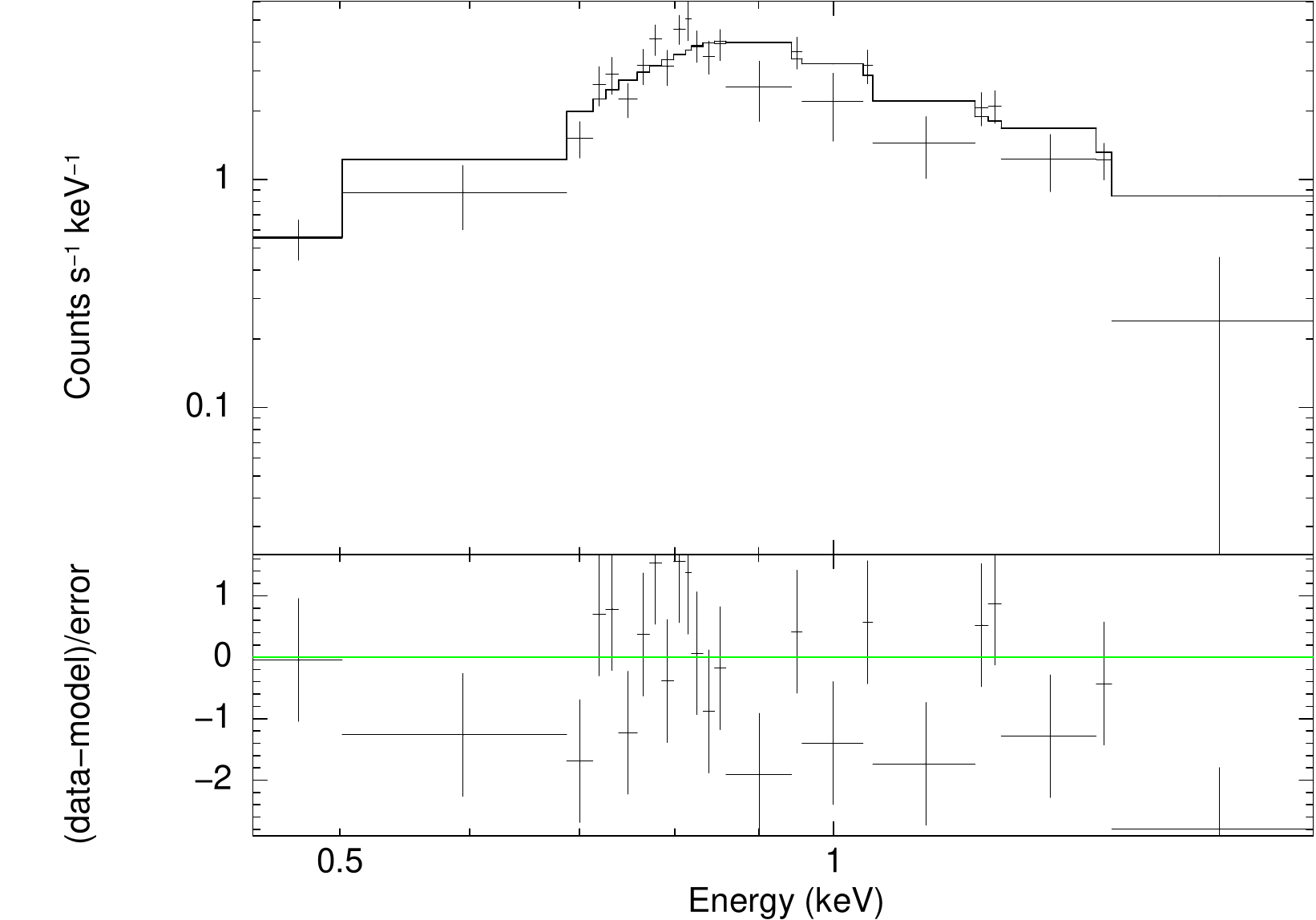}
    \caption{\textit{eROSITA} spectrum of the foreground source (blue elliptical region shown in Figure~\ref{fig:erositadr1}). Spectrum is fitted with the model \textit{TBabs}(\textit{powerlaw}+\textit{nei}).}
    \label{fig:foregroundspec}
\end{figure}

\begin{figure*}
	\centering
	\subfloat[vpshock]{\includegraphics[height=\columnwidth,angle=-90]{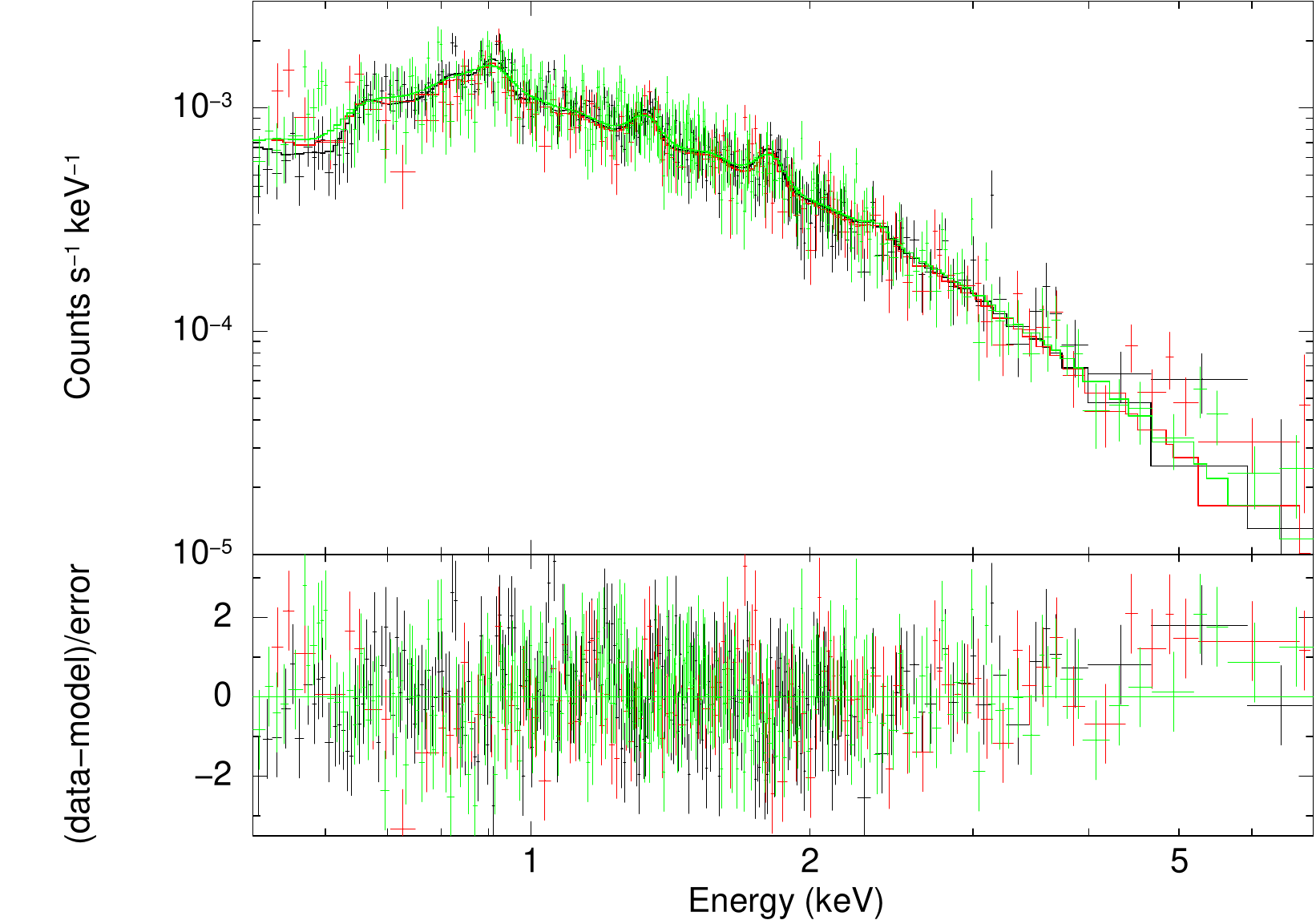}
		\label{fig:southvpshock}}
	\subfloat[vpshock+powerlaw]{\includegraphics[height=\columnwidth,angle=-90]{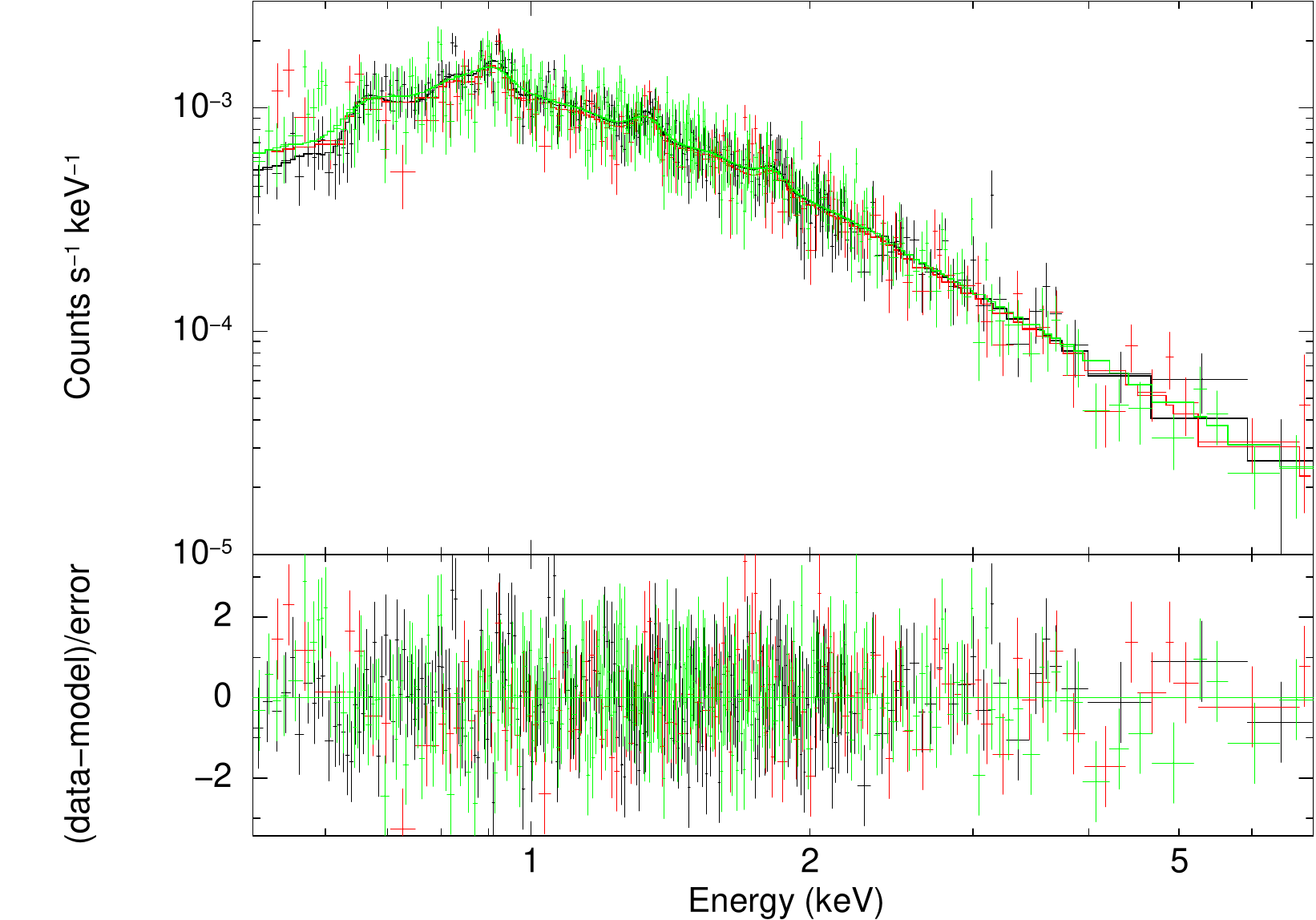}
		\label{fig:southvpshockpow}}
	\caption{\textit{eROSITA} and \textit{XMM-Newton} spectrum of the southern shell with different models (black: \textit{eROSITA} spectrum; red: MOS2 spectrum of \textit{XMM-Newton}; green: pn spectrum of \textit{XMM-Newton}.)}
	\label{fig:southshellspec}
\end{figure*}

Inspection of the \textit{eROSITA}-DE Data Release~1 image, together with \textit{ASKAP} radio data, reveals an additional X-ray source in the southeast of the SNR (Figure~\ref{fig:erositadr1}). Its morphology and spectral properties indicate that it is likely a candidate foreground SNR. The spectrum is best described by a combination of a \textit{powerlaw} and \redbf{\textit{nei} components}, yielding $N_{\mathrm{H}} \sim 0.37$~$\times$~10$^{22}$~cm$^{-2}$, photon index $\Gamma \sim 2.19$, and plasma temperature $kT \sim 0.78$~keV (Figure~\ref{fig:foregroundspec}). Because this source overlaps with the extraction region of the southern shell, it contaminates the shell's spectrum.

To account for this contamination, we scaled the foreground model by area and included it in the joint fit. After doing so, the southern shell spectra can be described either by a high-temperature plasma ($kT \sim 1.57~\mathrm{keV}$) or by a combination of a lower-temperature thermal component ($kT \sim 0.73~\mathrm{keV}$) and a \textit{powerlaw}, similar to the northern shell. The final fit including the foreground source is presented in Table~\ref{tab:shellspec}.
In \redbf{Section}~\ref{sec:properityofsnr}, we discuss the implied physical properties of the SNR.

\begin{table*}[htbp]
\centering
\caption{Best-fit results and 90\% uncertainties
{for the ``south'' and ``north'' shell regions shown in Figure \ref{fig:allregion} \redbf{after including}
a foreground component, 
using a two-component absorbed model (\textit{vpshock} for the thermal component and \textit{powerlaw} for the potential non-thermal component using \textit{XMM-Newton} and \textit{eROSITA}).}
}
\label{tab:shellspec}
\begin{tabular}{lcccc}
\hline\hline
 & south(\textit{vpshock}) & south(\textit{vpshock} + \textit{powerlaw}) & north(\textit{vpshock}) & north(\textit{vpshock} + \textit{powerlaw}) \\
\hline\hline
${N}_{\mathrm{H}}\ \rm{10^{22}\ cm^{-2}}$ & $0.82^{+0.05}_{-0.05}$ & $0.93^{+0.10}_{-0.10}$ & $0.92^{+0.12}_{-0.08}$ & $1.05^{+0.22}_{-0.14}$ \\
$\textit{kT\ \rm{(keV)}}$ & $1.57^{+0.14}_{-0.12}$ & $0.73^{+0.21}_{-0.18}$ & $3.41^{+0.47}_{-0.58}$ & $1.98^{+2.13}_{-0.71}$ \\
\textit{O} & $0.11^{+0.04}_{-0.03}$ & $0.20^{+0.17}_{-0.07}$ & $0.06^{+0.07}_{-0.04}$ & $0.17^{+0.35}_{-0.10}$ \\
\textit{Ne} & $0.12^{+0.03}_{-0.02}$ & $0.16^{+0.13}_{-0.05}$ & $0.19^{+0.05}_{-0.04}$ & $0.28^{+0.25}_{-0.09}$ \\
\textit{Mg} & $0.30^{+0.08}_{-0.08}$ & $0.36^{+0.33}_{-0.15}$ & $0.27^{+0.10}_{-0.10}$ & $0.32^{+0.12}_{-0.13}$ \\
$\tau$($10^{10}$  $\rm cm^{-3}\ s)$ & $0.36^{+0.03}_{-0.03}$ & $0.53^{+0.21}_{-0.09}$ & $0.32^{+0.04}_{-0.04}$ & $0.38^{+0.11}_{-0.07}$ \\
$\textit{norm}\ (10^{-3} \ \rm{cm^{-5}})$ & $8.2^{+0.8}_{-0.8}$ & $11.1^{+6.3}_{-6.7}$ & $4.8^{+0.6}_{-0.3}$ & $4.5^{+2.5}_{-2.5}$ \\
\textit{Photon Index} & - & $2.2^{+0.4}_{-0.5}$ & - & $1.5^{+1.6}_{-1.1}$ \\
$\textit{Photon norm ($10^{-3}$)}$ & - & $1.0^{+0.8}_{-0.6}$ & - & $0.4^{+0.6}_{-0.3}$ \\
$\chi^2_{\nu}$(d.o.f.) & 1.132(856) & 1.048(854) & 1.071(196) & 1.047(194) \\
\hline
\end{tabular}

\end{table*}

\section{Discussion}

\subsection{Properties of the SNR}
\label{sec:properityofsnr}

We adopted the pulsar position reported by \cite{Zhang_2025}, 
R.A. = 15h13m55.6s and Dec. = -$59^{\circ}08^{\prime}09.4^{\prime\prime}$ (J2000),
and assumed a distance of 5.2 kpc. {Taking the pulsar position as the explosion center, under the assumption that the pulsar has remained close to its birth site,}
we measured the angular radii of the northern and southern shells of the SNR to be $10^{\prime}$ and $16^{\prime}$, corresponding to physical radii of 15.1 pc and 24.2 pc, respectively. {Because of the pronounced asymmetry of the remnant, the northern and southern shell segments were analyzed independently. The derived dynamical parameters therefore characterize the local shell properties and should not be interpreted as indicating different evolutionary ages for different parts of the remnant.}

The shock velocity can be estimated using

\[
v_s = \sqrt{\frac{16 \, kT}{3 \, \mu \, m_p}},
\]
where \(m_p\) is the proton mass and \(\mu = 0.61\) is the mean mass per particle for a fully ionized plasma \citep{1967ZelDovich}. {We estimate the shock velocity assuming electron-ion temperature equilibration, following the standard approach commonly adopted in SNR studies. Since the post-shock proton temperature is generally expected to exceed the electron temperature when equilibration is incomplete, the derived shock velocity should be regarded as a lower limit.}

For the pure thermal model, the X-ray emission originates from the shocked plasma, and the measured temperatures of $3.41^{+0.47}_{-0.58}$ keV (north) and $1.57^{+0.14}_{-0.12}$ keV (south), 
yield shock velocities of {\(1700\ \pm 150\)} km/s and {\(1150\ \pm\ 50\)} km/s, respectively. 
Assuming the SNR is in the Sedov phase, the ages inferred from \(t = \frac{2}{5}\frac{R}{v_s}\) \citep[][]{1950Taylor,1950Taylor2, Sedov_1959, Truelove_1999}
(with R = 15.1 pc and R = 24.2 pc for the north and south shells) are \( 3.5\ \pm\ 0.3\) kyr and \( 8.3\ \pm\ 0.4\) kyr, respectively. 
{For the density estimates, we assumed an ellipsoidal geometry for each shell and took the semi-minor axis of the projected elliptical regions as the line-of-sight depth, corresponding to a thin-shell approximation along the viewing direction; the resulting volumes are}
$1.33\ \times\ 10^{57}\ \mathrm{cm^3}$ and $5.40\ \times\ 10^{57}\ \mathrm{cm^3}$ for the north and south shells, respectively. The \textit{XSPEC} thermal plasma normalization is defined as
\[
\mathrm{Norm} = \frac{10^{-14}}{4\pi d^{2}} \int n_{\mathrm{e}}\, n_{\mathrm{H}}\, dV,
\]
where \(d = 5.2\ kpc\) is the source distance, and \(n_{\mathrm{e}}\) and \(n_{\mathrm{H}}\) are the electron and hydrogen number densities \citep{2001Borkowski}. For an ionized plasma with solar or slightly enhanced metal abundances, we assume ${n}_\mathrm{e}=1.2{n}_\mathrm{H}$. In the pure thermal case, the derived hydrogen number densities are $0.98\pm0.06\ \mathrm{cm}^\mathrm{-3}$ for the northern shell and $0.64\pm0.03\ \mathrm{cm}^\mathrm{-3}$ for the southern shell. The upstream ambient density is then estimated as $n_0=n_H/4$ for a strong shock. Using the Sedov relation for the explosion energy, 

\[
E_0=25(1.4{n_0}{m_H})R^3_sV_s^2/(4\xi)
\]
 with $\xi\ =\ 2.026$ for $\gamma=5/3$ \citep[][]{1950Taylor,1950Taylor2, Sedov_1959, Truelove_1999} we obtain explosion energies of $(5.1 \ \pm\ 0.9)\times10^{51}$ 
erg and $(6.3 \ \pm\ 0.6)\times10^{51}$ erg
for the northern and southern shells, respectively, in the single-component case. We note that if the electron temperature is lower than the ion temperature, the derived shock velocities, ages, and explosion energies should be regarded as upper limits. Under such conditions, the estimated ages of the northern (3.5 kyr) and southern (8.3 kyr) shells are 
larger than the pulsar's characteristic age of 1.7 kyr \citep[][]{kaspi_1994,Gaensler_1999,Ye_2025}.

For the \textit{vpshock} model with an additional \textit{powerlaw} component, the best-fit temperatures are lower than those in the pure \textit{vpshock} case. For the north and south shells, the temperatures are $1.98^{+2.13}_{-0.71}\ \rm{keV}$ and $0.73^{+0.21}_{-0.18}\ \rm{keV}$, respectively. 
Using the same method as above, we obtain shock velocities of  {$1300\ \pm\ 700$} km/s and {$800\ \pm\ 100$} km/s, 
and Sedov ages of $4.6\ \pm\ 2.5\ \rm{kyr}$ and $12.1\ \pm\ 1.7\ \rm{kyr}$ for the north and south shells, respectively. 

\begin{figure}[htbp]
    \centering    \includegraphics[width=\columnwidth]{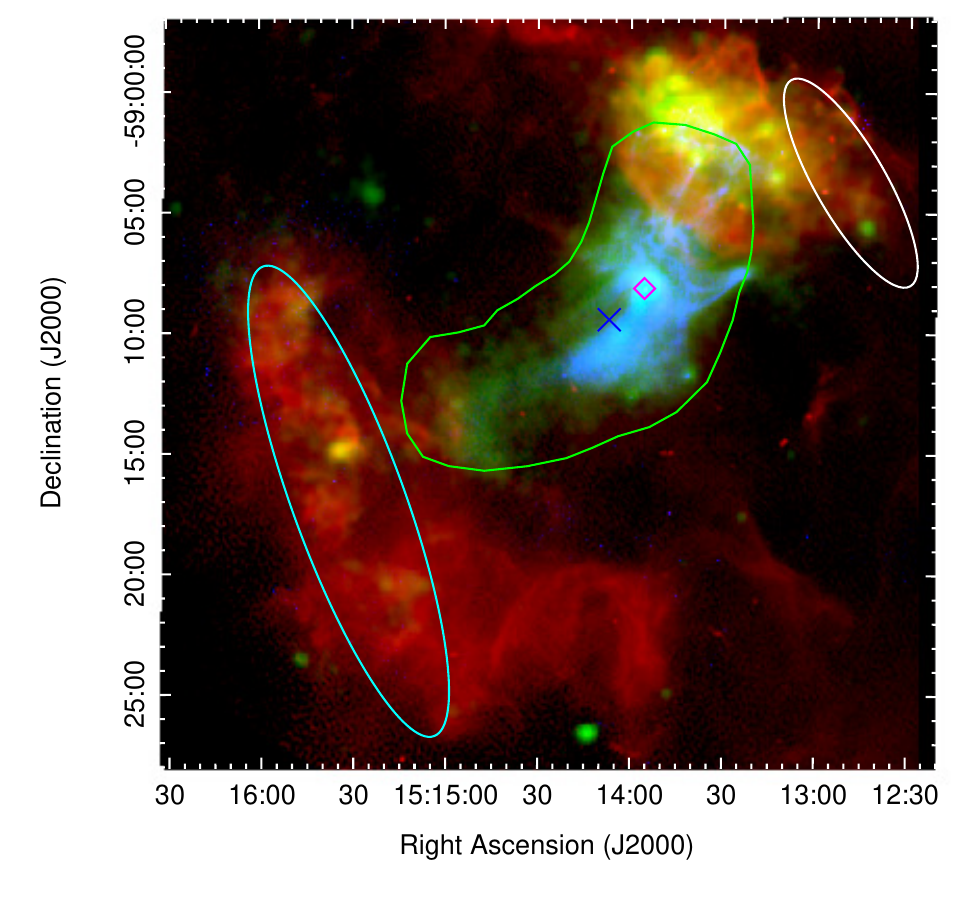}
    \caption{Multi-wavelength image of MSH~15--52 (red: \textit{MeerKAT} image, green: 500--2000 eV \textit{eROSITA} image, blue: 3000--8000 eV \textit{Chandra} image). Contour of H.E.S.S. 30 excess map shown in green polygon \citep{Aharonian_2005}. The blue cross indicates the centroid of the HESS $\gamma$-ray emission, which was derived by fitting the HESS data with a Gaussian profile \citep{Aharonian_2005}. The magenta diamond indicates the position of the pulsar, PSR~J1513--5908.}
    \label{fig:mutiwavelengthimg}
\end{figure}

In this two-component scenario, the corresponding hydrogen number densities are $0.95\pm0.27\ \mathrm{cm}^\mathrm{-3}$ for the northern shell and $0.74\pm0.19\ \mathrm{cm}^\mathrm{-3}$ for the southern shell. The inferred explosion energies are $(2.9 \ \pm\ 3.2)\times10^{51}$~erg and $(3.4 \ \pm\ 1.3)\times10^{51}$~erg for the northern and southern shells, in the \textit{vpshock+powerlaw} case.
The photon index of the \textit{powerlaw} component in the shell region is $2.16^{+0.37}_{-0.52}$ in the south and $1.52^{+1.61}_{-1.10}$ in the north, indicating a relatively hard non-thermal spectrum. Such spectral hardening suggests that the shell emission may be influenced by particle injection from the PWN jets, similar to what has been proposed for the W50 nebula \citep{Safi-Harb_2022}. This scenario is further supported by the low ionization timescale inferred from the thermal fit, suggesting recently shocked plasma. As shown in Figure~\ref{fig:mutiwavelengthimg}, the gamma-ray emission also extends into the shell, further supporting a possible connection between the PWN outflow and the shell region. Moreover, the photon index we measure in the northern and southern parts of the shell is consistent, within uncertainties, with the outermost photon index ($2.24 \pm 0.28$) reported in the concentric-ring analysis of the PWN \citep{2010_schock}.

\subsection{Properties of the PWN jets}

We investigated the potential northern jet region, which could align with the pointer or the thrid finger of the ``Hand of God." Extracting and modeling 6 regions of similar surface brightness, we expect harder photon indices to trace the path of the jet. The resulting photon index map, seen in Figure \ref{fig:photonIndex_pants}, indicates a relatively hard photon index that softens with distance from the pulsar. The regions 0, 3, and 5, which trace the ``pointer finger" has a slightly lower photon index throughout, which suggests a more likely path for the jet. \cite{Romani_2023} analyze each finger individually and also report that the pointer finger has an overall harder photon index than the third finger. 

A slightly unexpected result was the abrupt hard-to-soft photon indices detected from regions 4 to 5 of the southern jet as seen in Figure \ref{fig:photonIndex_southJet}. Region 4 is fit with the hardest photon index and corresponds to the location of a surface brightness change from opaque to more diffuse in the merged image (Figure \ref{fig:Chandra_mapRegs}). This coincidence suggests an interaction with the reverse shock at that location, which would explain the very hard photon index and the sudden softening.
{However, establishing a reverse-shock interaction requires comparison with dedicated PWN–SNR simulations \citep{Temim_2017, Kolb_2017}, as morphology alone cannot uniquely diagnose such an interaction. This is beyond the scope of this paper and will be addressed in future work. However, to further investigate this scenario}, we consider the characteristic timescale for the collision between the PWN and the reverse shock as discussed in \citet{vanderSwaluw_2004}. Using their expression,

\[
t_{\mathrm{col}} = 1045\, E_{51}^{-1/2}
\left( \frac{M_{\mathrm{ej}}}{M_{\odot}} \right)^{5/6}
n_0^{-1/3} \ \mathrm{yr}
\]

where $E_{51}$ is the explosion energy in units of $10^{51}\,\mathrm{erg}$ and $M_{\mathrm{ej}}$ is the ejecta mass, we estimate the collision timescale by adopting $M_{\mathrm{ej}} = 10\,M_{\odot}$. The explosion energy and ambient density used here are taken from the analysis in Section~\ref{sec:properityofsnr}. Because the shell properties were derived separately for the northern and southern regions, and each region was modeled with both single- and double-component fits, four collision timescales are obtained: $5.0 \pm 0.5$~kyr and $6.8 \pm 3.8$~kyr for the northern and southern shells in the single-component model, and $5.2 \pm 0.3$~kyr and $6.8 \pm 1.4$~kyr in the double-component model, respectively. These values are broadly comparable to the Sedov ages derived earlier, suggesting that interaction between the reverse shock and the PWN is likely occurring.

However, these estimates rely on several assumptions and should therefore be interpreted with caution. For example, the shock velocity is inferred from the electron temperature, although electrons may not be fully equilibrated with ions immediately behind the shock. The ejecta mass is also assumed rather than directly constrained, and part of the shell energy may originate from the jets rather than solely from the supernova blast wave. Given these uncertainties, the collision timescales are used here primarily as a qualitative indicator of the presence of reverse-shock--PWN interaction.

{\cite{An_2014} investigated spatial variations in the photon index in detail while also allowing the column density ($N_H$) to vary. They found that the variations in the photon index were substantially reduced when $N_H$ was allowed to vary. We find a similar behavior in our analysis of the southern jet. Although allowing $N_H$ to vary improves the reduced $\chi^2$ for some regions, such large variations in $N_H$ are not expected on the small spatial scales considered here. We therefore adopt a fixed $N_H$ for the jet regions, which is more physically motivated if the regions trace emission from the same structure. The free-$N_H$ fits are provided in Appendix Table \ref{tab:alljetregionfit_freenH} for completeness. The main trends in the photon index remain when $N_H$ is allowed to vary, although the differences between regions are less pronounced.}

\subsection{Supernova Progenitor of MSH~15--52}

\label{subsec:progenitor}

To deduce the properties of the progenitor of the SNR MSH~15--52, 
{we compared the observed elemental abundances of its ejecta clumps (``ejecta1'',``ejecta2'', ``ejecta3'', and ``ejecta4'') with theoretical nucleosynthesis yields from SN explosion models of massive stars.}
Here we use two approaches:  first, we compare total supernova yields from one-dimensional models to our observations; second, we match each observed region to regions of an asymmetric explosion.  

Comparing total supernova yields places strong limits on the supernova progenitor.  For this comparison, we use theoretical results from the systematic survey of \cite{Sukhbold2016}. In that work, the authors followed the complete evolution and explosion of massive stars with zero-age main sequence (ZAMS) masses from 9 $\rm{M}\odot$ to 120 $\rm{M}\odot$, calculating detailed nucleosynthesis based on a neutrino-powered explosion mechanism.

\begin{figure*}[htbp]
    \centering

    \subfloat[ejecta1]{%
        \includegraphics[width=0.45\textwidth]{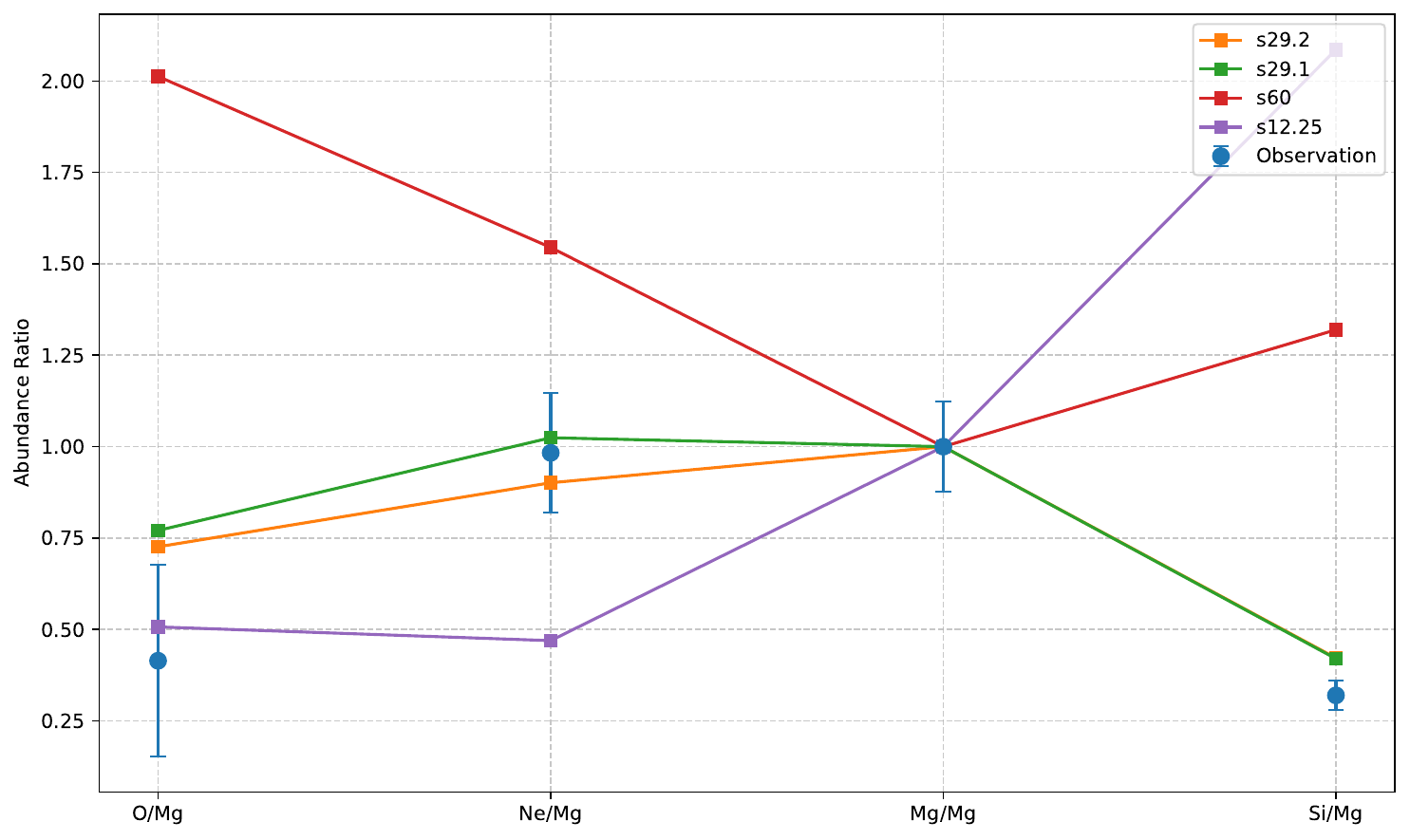}%
        \label{fig:ejecta1compare}
    }
    \hfill
    \subfloat[ejecta2]{%
        \includegraphics[width=0.45\textwidth]{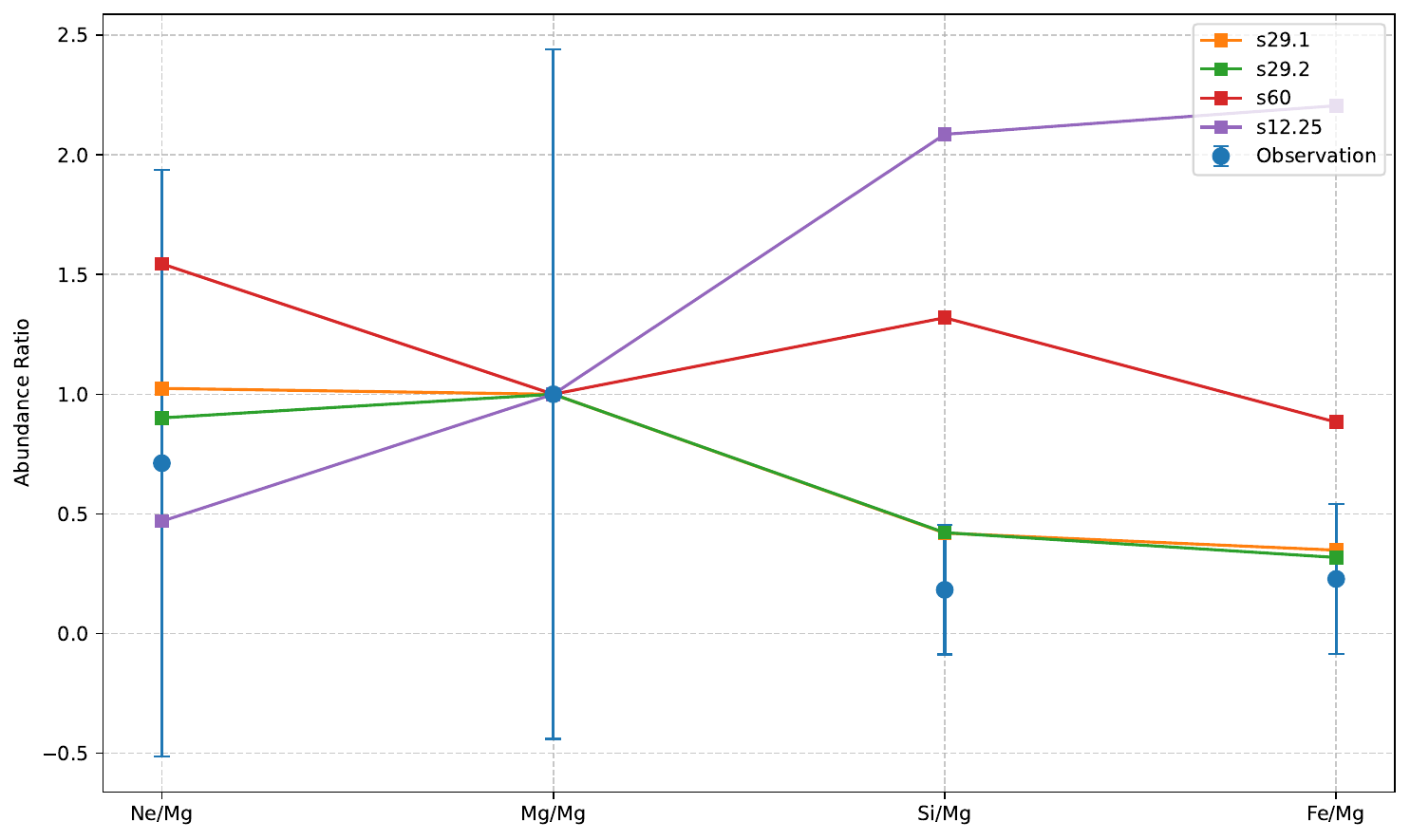}%
        \label{fig:ejecta2compare}
    }

    \subfloat[ejecta3]{%
        \includegraphics[width=0.45\textwidth]{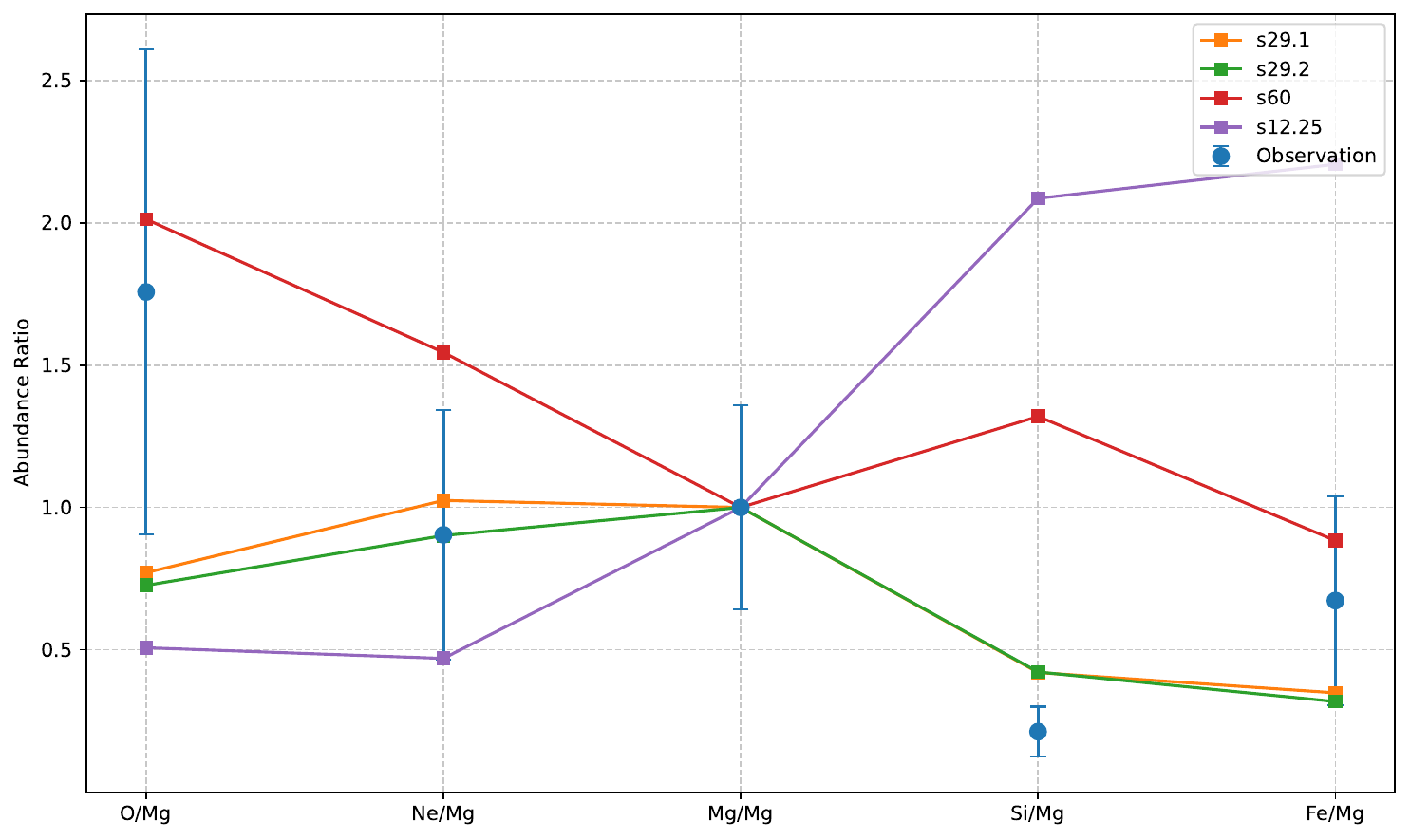}%
        \label{fig:ejecta3compare}
    }
    \hfill
    \subfloat[ejecta4]{%
        \includegraphics[width=0.45\textwidth]{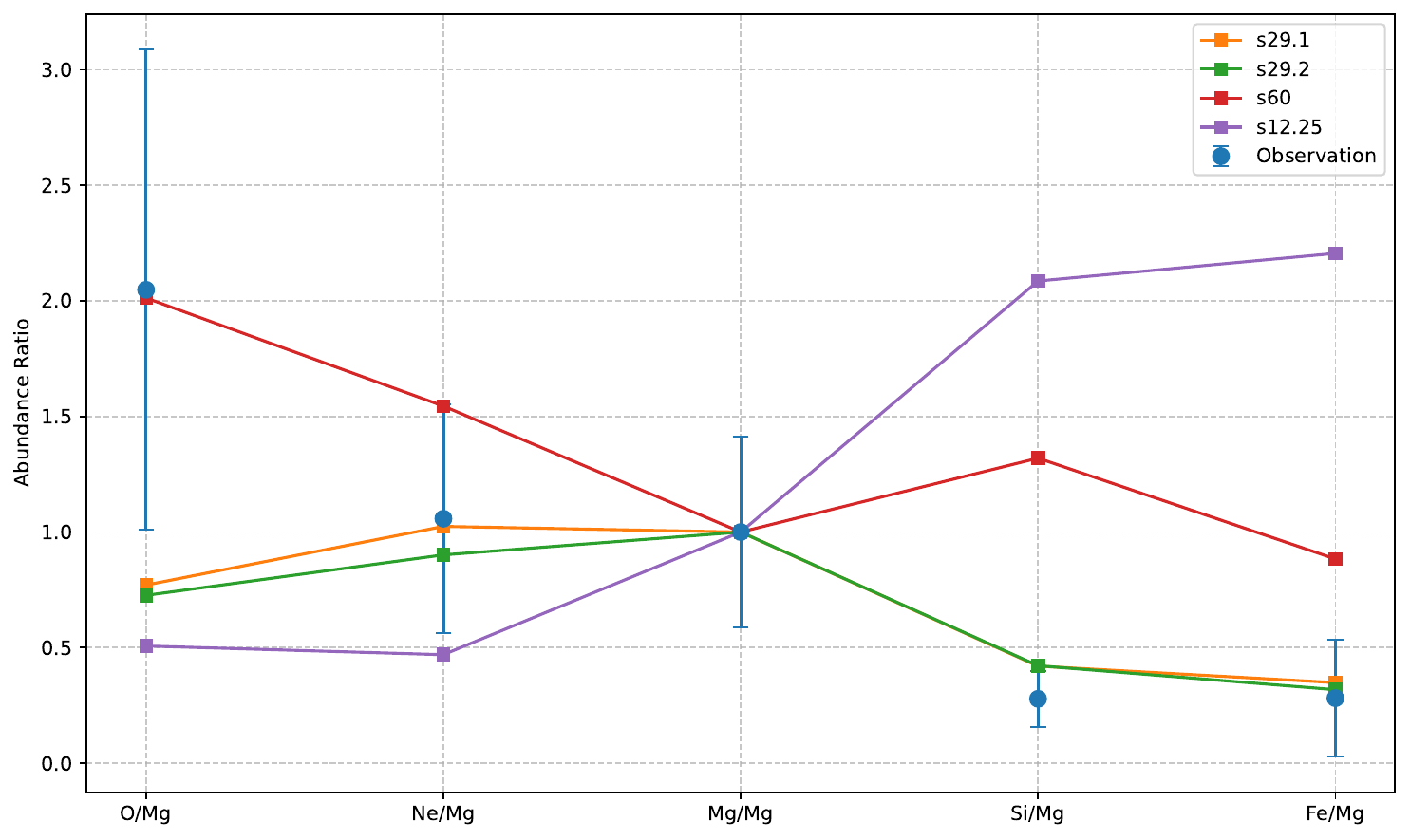}%
        \label{fig:ejecta4compare}
    }

    \caption{Comparison between the abundance ratios of the ejecta regions and the nucleosynthesis models of \cite{Sukhbold2016}.}
    \label{fig:comparsukhbold}
\end{figure*}

\cite{Sukhbold2016} used several “central engines”, parameterized models for neutrino emission from the inner proto-neutron star to reproduce the observed properties of SN~1987A and SN~1054. Among the different engines tested, only the ones calibrated to the SN~1987A progenitor models W18 and N20 yielded a sufficient number of successful explosions and an integrated nucleosynthesis output broadly consistent with solar abundances. Therefore, the results from these two engines are considered the most physically reliable for comparison with observations.

In this study, we adopt the nucleosynthesis yields predicted by the N20 central engine for our analysis. As shown in Figure \ref{fig:comparsukhbold}, the abundance pattern measured in the
{``ejecta''}
region is best reproduced by the model corresponding to a progenitor with a ZAMS mass of 29.1 $\rm{M}_\odot$ ($\chi^2_{\nu}\sim1.28)$. Similar comparisons for the other regions {considered as ejecta} also favor progenitor masses of 29.1 $\rm{M}_\odot$ or 29.2 $\rm{M}_\odot$. These results consistently point to a massive star of mass $\sim29\ \rm{M_\odot}$ as the progenitor of this SNR.
For comparison, we also tested the W18 central engine. However, its predicted yields provide a significantly poorer match to the observed abundances ($\chi^2_{\nu}\sim3.03$).

But comparing the inferred total yields to one-dimensional models may be too simplified.  We should bear in mind that our observations only probe regions of the supernova ejecta.  It is likely that the different ejecta observations probe different regions of the supernova ejecta.  And these different regions can arise from different shock velocities caused by the fact that the convection-enhanced supernova engine produces asymmetric explosions.  As such, by studying the different observations, we can probe the extent of the asymmetries.  

\begin{figure}[htbp]
    \centering

    \includegraphics[width=\columnwidth]{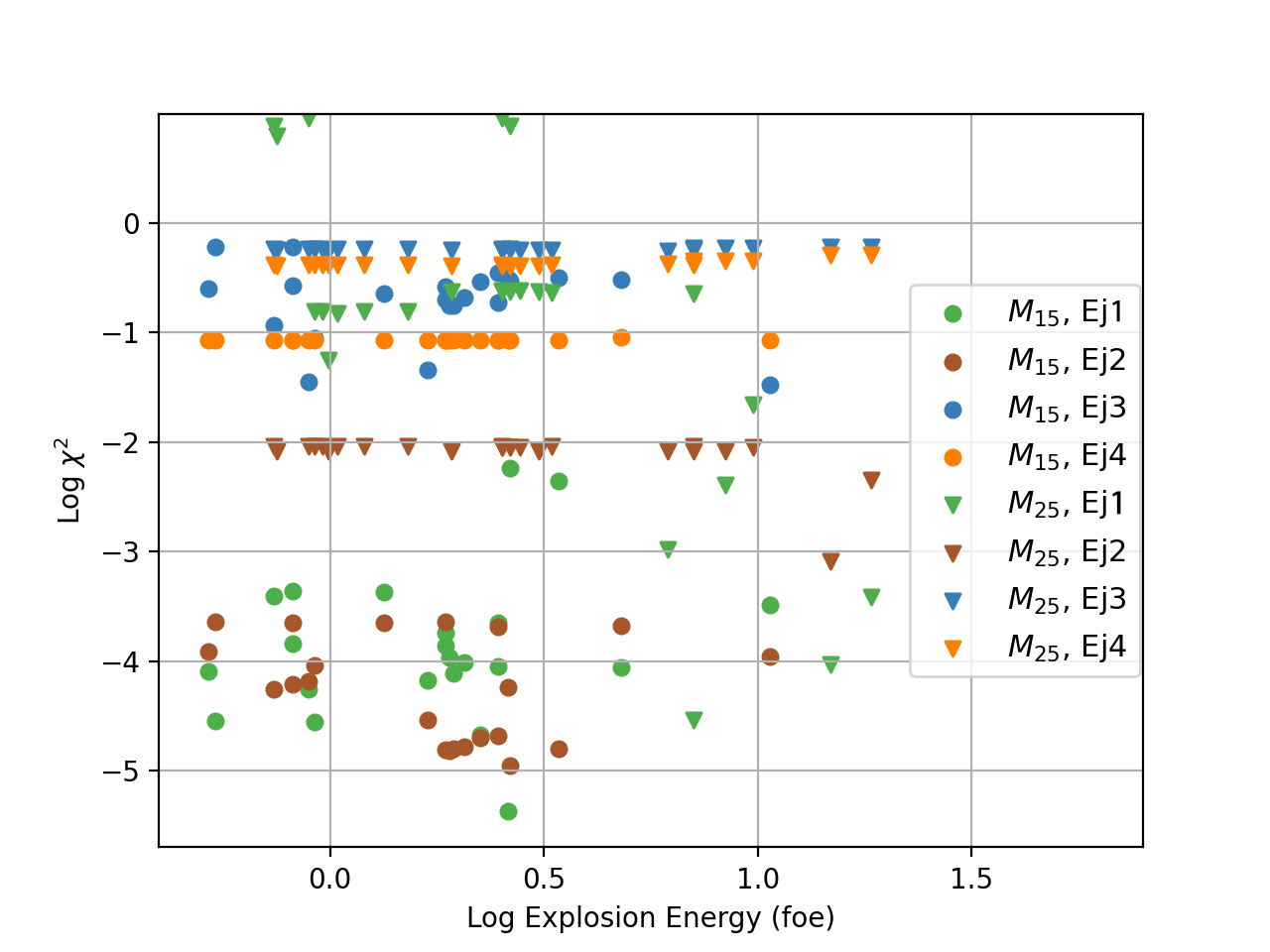}
    \caption{Log $\chi^2_{\nu}$ fits for our 4 different ejecta regions versus energy for 15 and 25\,M$_\odot$ progenitors with a range of explosion energies.  We use 2 separate regions from the broad energy explosions from \cite{2018ApJ...856...63F,2020ApJ...890...35A}.  Different regions are best fit by different explosion energies.}
    \label{fig:chi2fit}
\end{figure}

Asymmetric explosions have been discussed in various contexts in the literature \citep{2025soker}, although without quantitative predictions that can be directly compared to our data.  Here, we model the asymmetries using the suite of explosion energies and yields from \cite{2018ApJ...856...63F,2020ApJ...890...35A}.
We assume that each of our observations is probing only a subset of the ejecta (i.e. a limited region in mass from the explosion).  To incorporate mixing in the explosion, we combine 2 mass regions.  We vary the extent and location of these mass regions and compare to each of our observations.  Figure~\ref{fig:chi2fit} shows the $\chi^2_{\nu}$ fits to the 4 observations for a range of explosion energies and 2 different progenitor masses.  The 15\,M$_\odot$ tends to produce slightly better fits, but both progenitors have ejecta that fit the observed data for some explosion energies with $\chi^2_{\nu}<\sim 0.1$.

\begin{figure*}[htbp]
    \centering
    \subfloat[ejecta1]{%
        \includegraphics[width=0.48\textwidth]{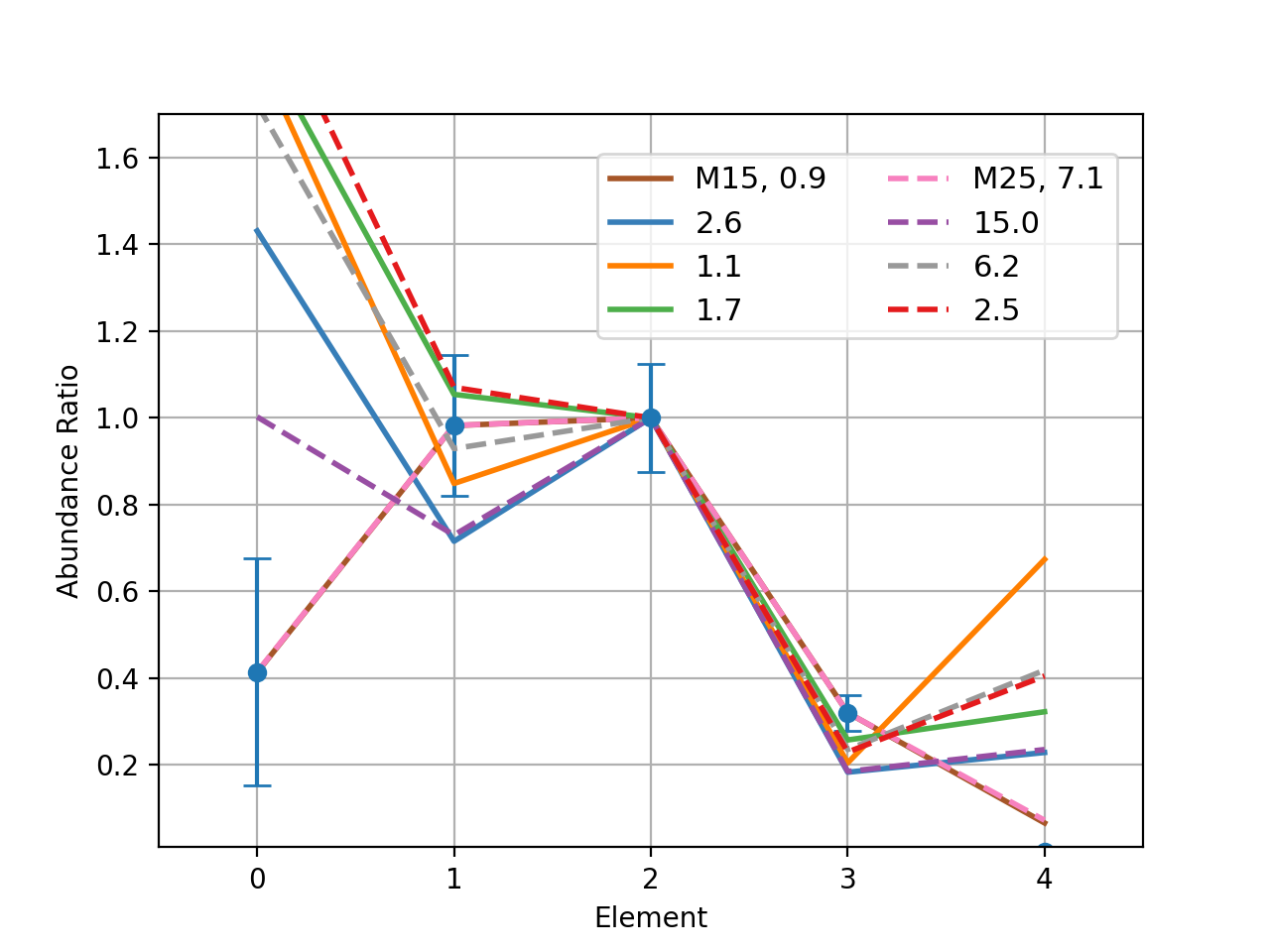}
        \label{fig:fobs1}
    }
    \subfloat[ejecta2]{%
        \includegraphics[width=0.48\textwidth]{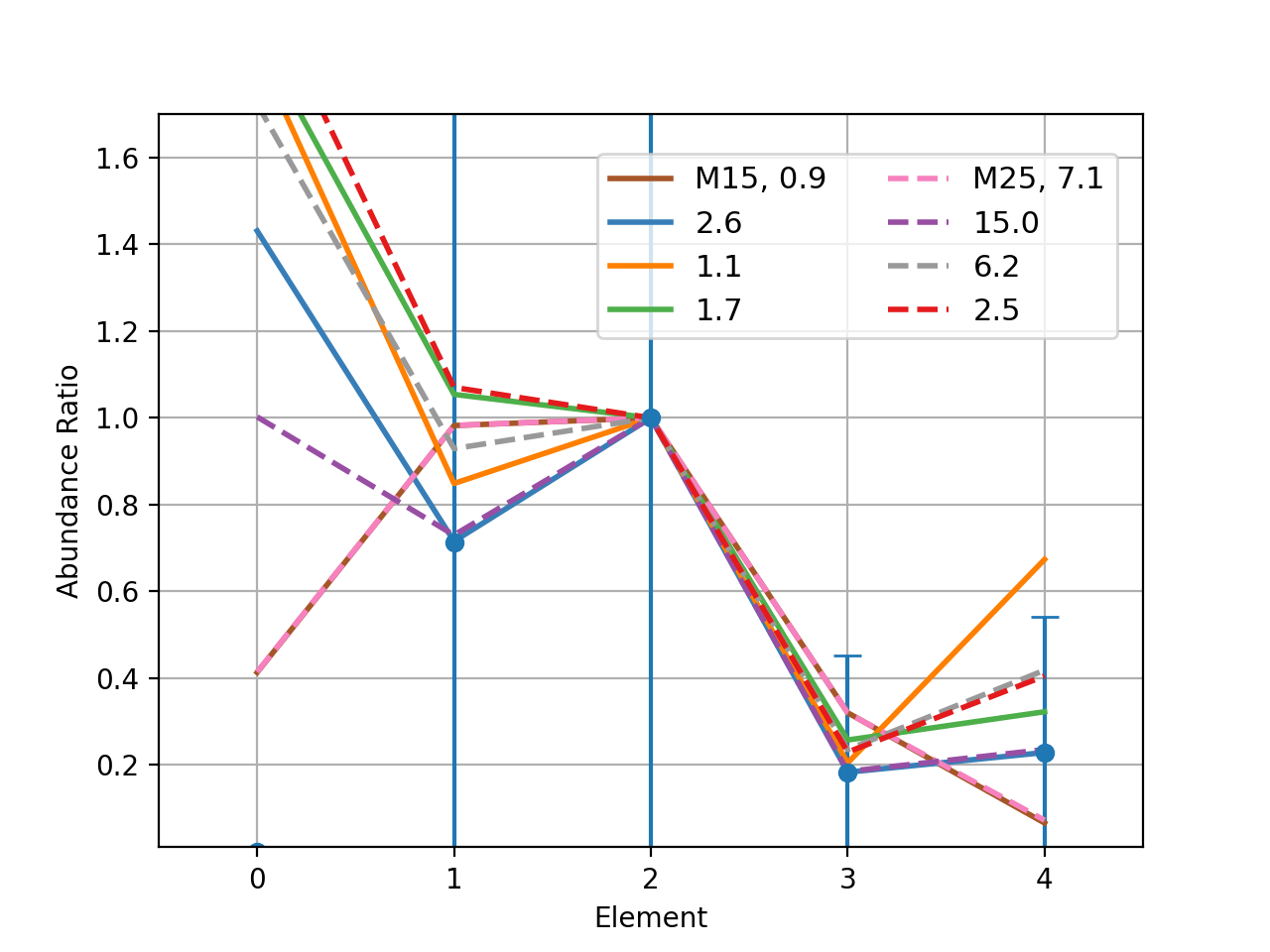}
        \label{fig:fobs2}
    }
    \\
    \subfloat[ejecta3]{%
        \includegraphics[width=0.48\textwidth]{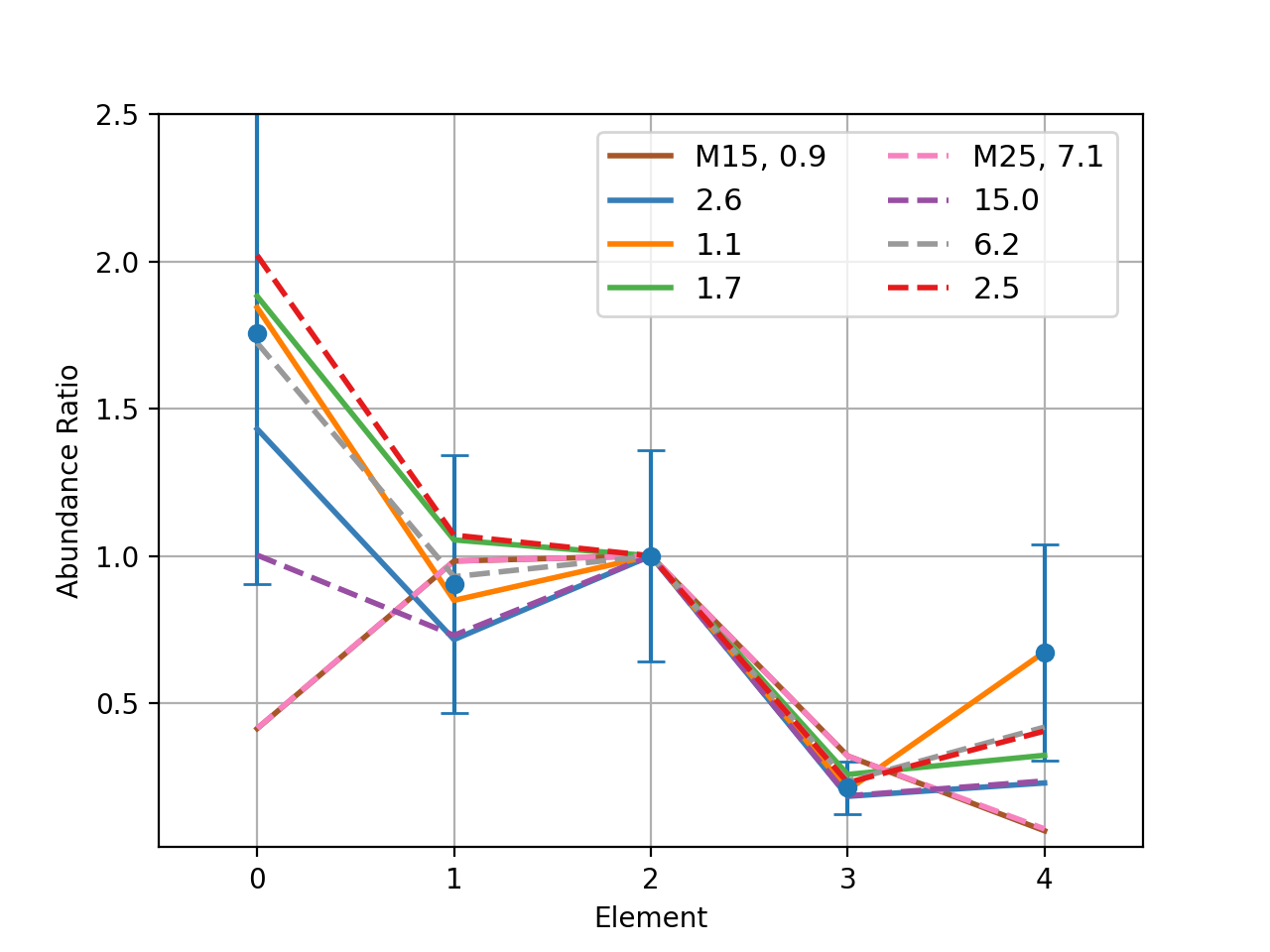}
        \label{fig:fobs3}
    }
    \subfloat[ejecta4]{%
        \includegraphics[width=0.48\textwidth]{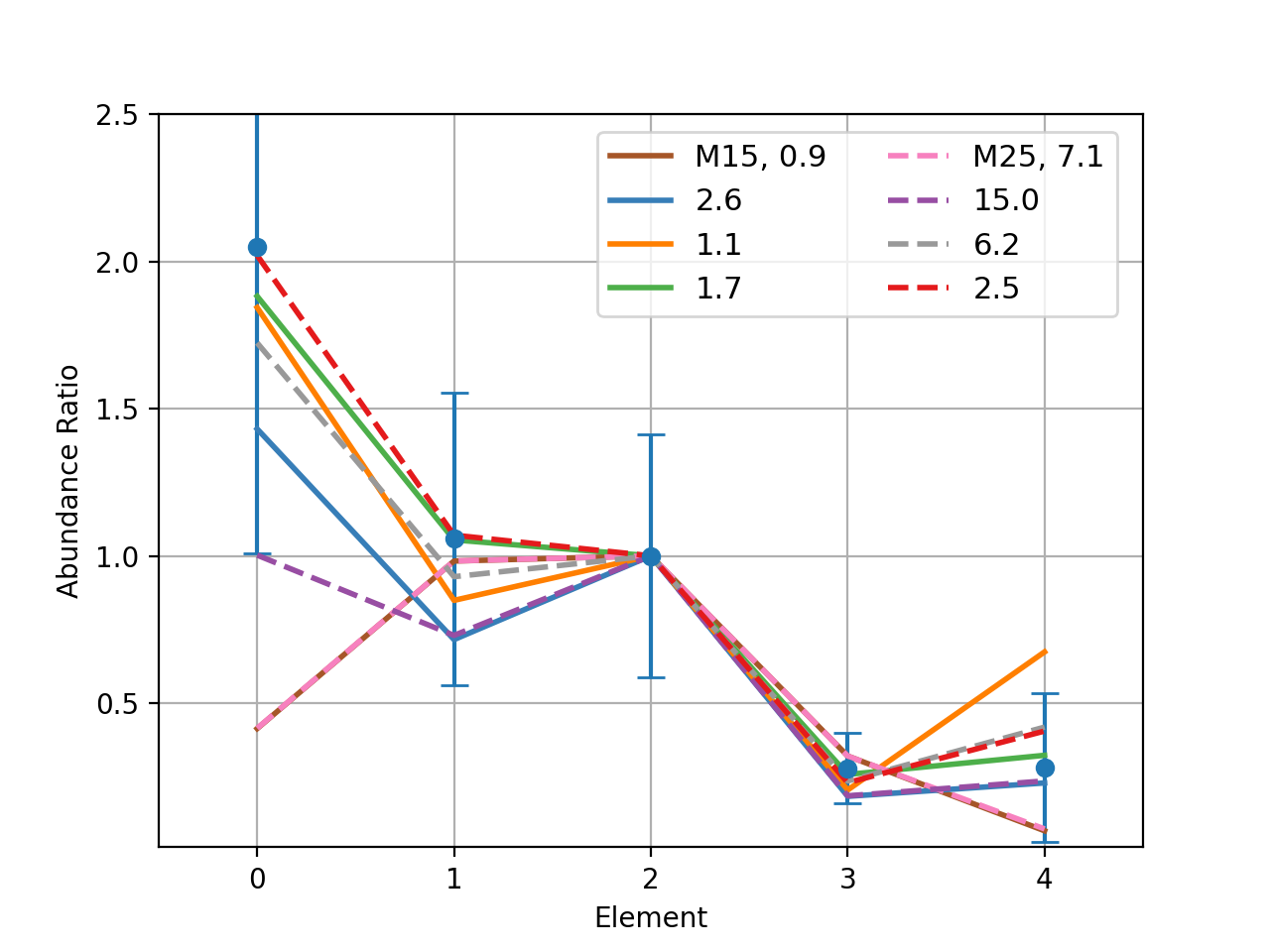}
        \label{fig:fobs4}
    }

    \caption{Abundance ratios versus element for our four observed ejecta regions with the best fit model for each individual observation on every plot.  
    The numbers correspond to elements: 0--oxygen, 1--neon, 2--magnesium, 3--silicon, 4--iron.  
    We include the best fits for the 15 and 25\,M$_\odot$ progenitors.  
    The best fit for any individual observation corresponds to a different energy.}
    \label{fig:fitasym}
\end{figure*}

For each progenitor mass, the best fit model for each observation corresponds to a different explosion energy (or a slightly different shock from an asymmetric supernova explosion).  Figure~\ref{fig:fitasym} shows the best fits of the abundance ratios to each observation.  The best fits for the 15\,M$_\odot$ model correspond to energies ranging from $0.9-2.6 \times 10^{51} {\rm erg}$.  The best fits for the 25\,M$_\odot$ model correspond to energies ranging from $2.5-15 \times 10^{51} {\rm erg}$.  Note that the best fit for one ejecta clump is often not the best fit for another clump.   

On the surface, this broad range of best-fit energies provides insight into the asymmetry in the explosion.  However, suitable fits can occur over a wide range of explosion energies with the current observational uncertainties.  With the current data, we cannot demonstrate wide asymmetries are needed, but as the data and simulations improve, data/simulation comparisons can place constraints on the explosion asymmetries.

\section{Conclusions}

In this paper, we analyzed X-ray data from \textit{XMM-Newton}, \textit{Chandra}, and \textit{eROSITA}. We presented X-ray images at different scales, identified new shell structures in both the northern and southern regions, and serendipitously discovered a new diffuse extended source (potentially a new foreground SNR). By comparing the abundances with nucleosynthesis models, we constrained the progenitor mass of MSH~15--52. The main conclusions are summarized as follows:

\begin{enumerate}
    \item 
    Based on the \textit{Chandra} observations, we detected strong emission lines in the spectrum of RCW 89. Spectral fitting indicates over-solar abundances of O, Ne, Mg, and Fe in ejecta clumps within RCW~89. 
    Comparison with core-collapse nucleosynthesis models favors
a massive progenitor with an initial mass of approximately $29\,M_\odot$ in one-dimensional yield models,
while comparisons with asymmetric explosion models show that lower-mass progenitors can also reproduce the observed abundance ratio.

    \item 
    Using the \textit{eROSITA} spectrum, we present the first analysis of the shell X-ray properties of this SNR. The spectrum can be fit either with a high-temperature thermal model or with a low-temperature thermal model combined with a non-thermal component. In both scenarios, the inferred Sedov age of the SNR is larger than the characteristic age of the pulsar, although the values remain model-dependent and subject to uncertainties in electron-ion equilibration, geometry, and shell contamination. In the second scenario, the presence of a very hard spectrum implies possible jet injection from the PWN and interaction with the SNR's shell component. Further and deeper observations are required to distinguish between these scenarios.

    \item 
    Using \textit{Chandra} data, we created new photon index maps of the southern and northern pulsar jet regions. We identify a probable path of the northern jet from pulsar B1509--58 to RCW 89. In the southern jet, we present evidence for a hard photon index region ($\Gamma$$\sim$1.6) 
    suggesting an interaction with the SNR reverse shock at this location.

    \item
    We identified a foreground source located southeast of the SNR. Spectral analysis together with the morphology observed in radio images suggests that this source is likely a foreground SNR. Additional deep observations are needed to confirm the nature of this source.
\end{enumerate}

\section*{Acknowledgments}

L.D. acknowledges support from the China Scholarship Council through the joint PhD training program between Nanjing University and the University of Manitoba, where this work was carried out.
I.S. acknowledges the support of a University of Manitoba Graduate Fellowship. S.S.H. acknowledges the support of the Natural Sciences and Engineering Research Council of Canada (NSERC) through the Discovery Grants Program and the Canada Research Chairs program, as well as the Canadian Space Agency.
We would like to thank Tianxian Luo for his valuable assistance with the radio data. We are also grateful for the insightful discussions with Ping Zhou and Yiheng Chi. 
{We are grateful to the referee for their thorough reading of the manuscript which helped improve the clarify and presentation of our results.}

This research made use of the NASA Astrophysics Data System (ADS), SNRcat -- the High-Energy Catalog of SNRs maintained at the University of Manitoba \cite{2012AdSpR..49.1313F}, and of HEASARC -- the High-Energy Astrophysics Science Archive Research Center maintained at the Goddard Space Flight Center. 
This work is based on data from \textit{eROSITA}, the soft X-ray instrument aboard SRG, a joint Russian-German science mission supported by the Russian Space Agency (Roskosmos), in the interests of the Russian Academy of Sciences represented by its Space Research Institute (IKI), and the Deutsches Zentrum für Luft- und Raumfahrt (DLR). The SRG spacecraft was built by Lavochkin Association (NPOL) and its subcontractors, and is operated by NPOL with support from the Max Planck Institute for Extraterrestrial Physics (MPE). The development and construction of the \textit{eROSITA} X-ray instrument was led by MPE, with contributions from the Dr. Karl Remeis Observatory Bamberg \& ECAP (FAU Erlangen-Nuernberg), the University of Hamburg Observatory, the Leibniz Institute for Astrophysics Potsdam (AIP), and the Institute for Astronomy and Astrophysics of the University of Tübingen, with the support of DLR and the Max Planck Society. The Argelander Institute for Astronomy of the University of Bonn and the Ludwig Maximilians Universität Munich also participated in the science preparation for \textit{eROSITA}. The \textit{eROSITA} data shown here were processed using the eSASS software system developed by the German \textit{eROSITA} consortium.

    \textit{Software:} DS9 \citep{2003Ds9_Joye}, SAS \citep{2014SAS}, CIAO \citep{2006Fruscione}, eSASS \citep[][]{esass_brunner_2018, 2021_Predehl_esass},  \textit{XSPEC} \citep{1996xspec_Arnaud}.

\clearpage

\appendix

\begin{figure}[htbp]
	\centering
	\subfloat[ejecta2]{\includegraphics[height=0.48\columnwidth,angle=-90]{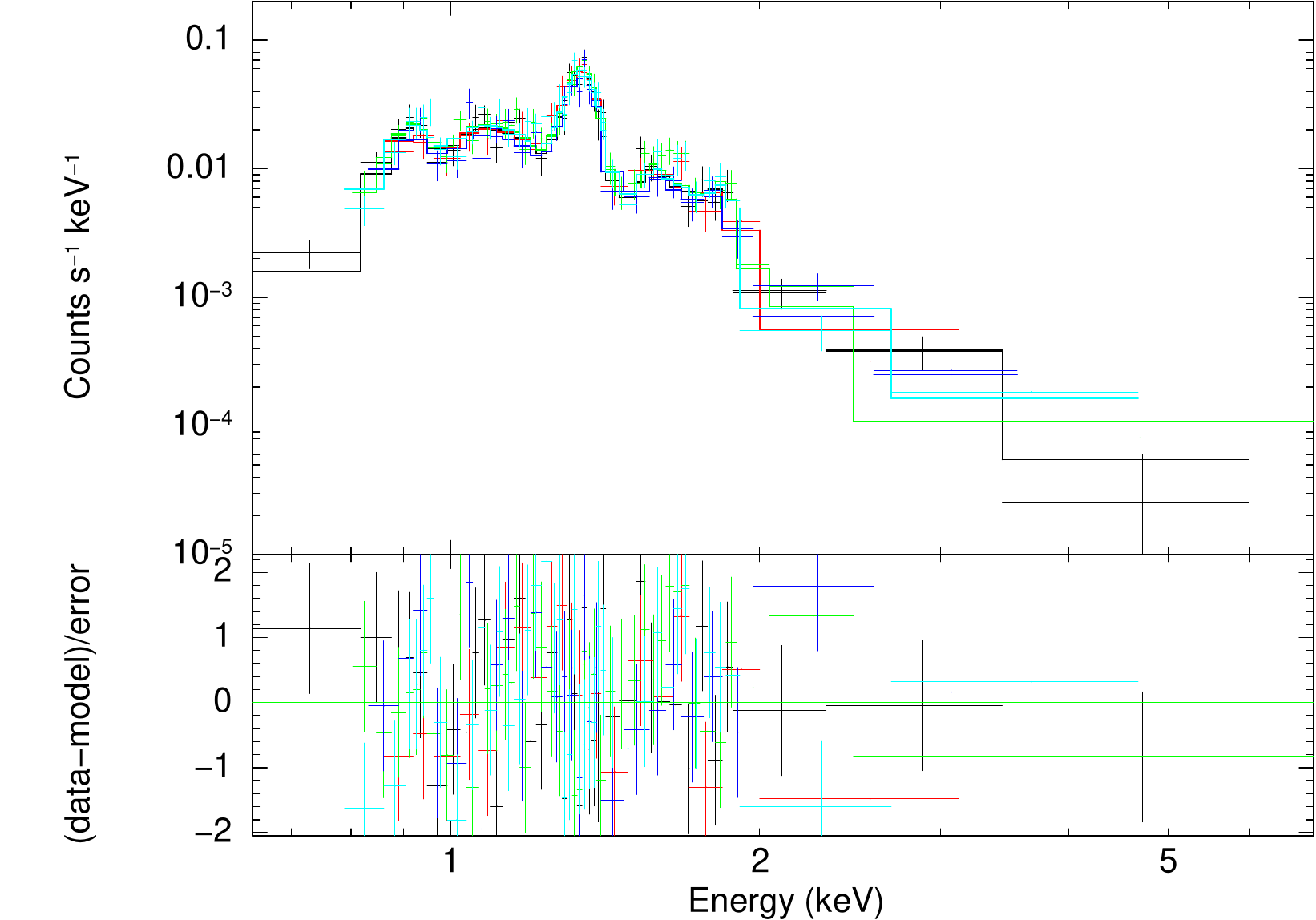}
		\label{fig:ejecta2spec}}
	\subfloat[ejecta3]{\includegraphics[height=0.48\columnwidth,angle=-90]{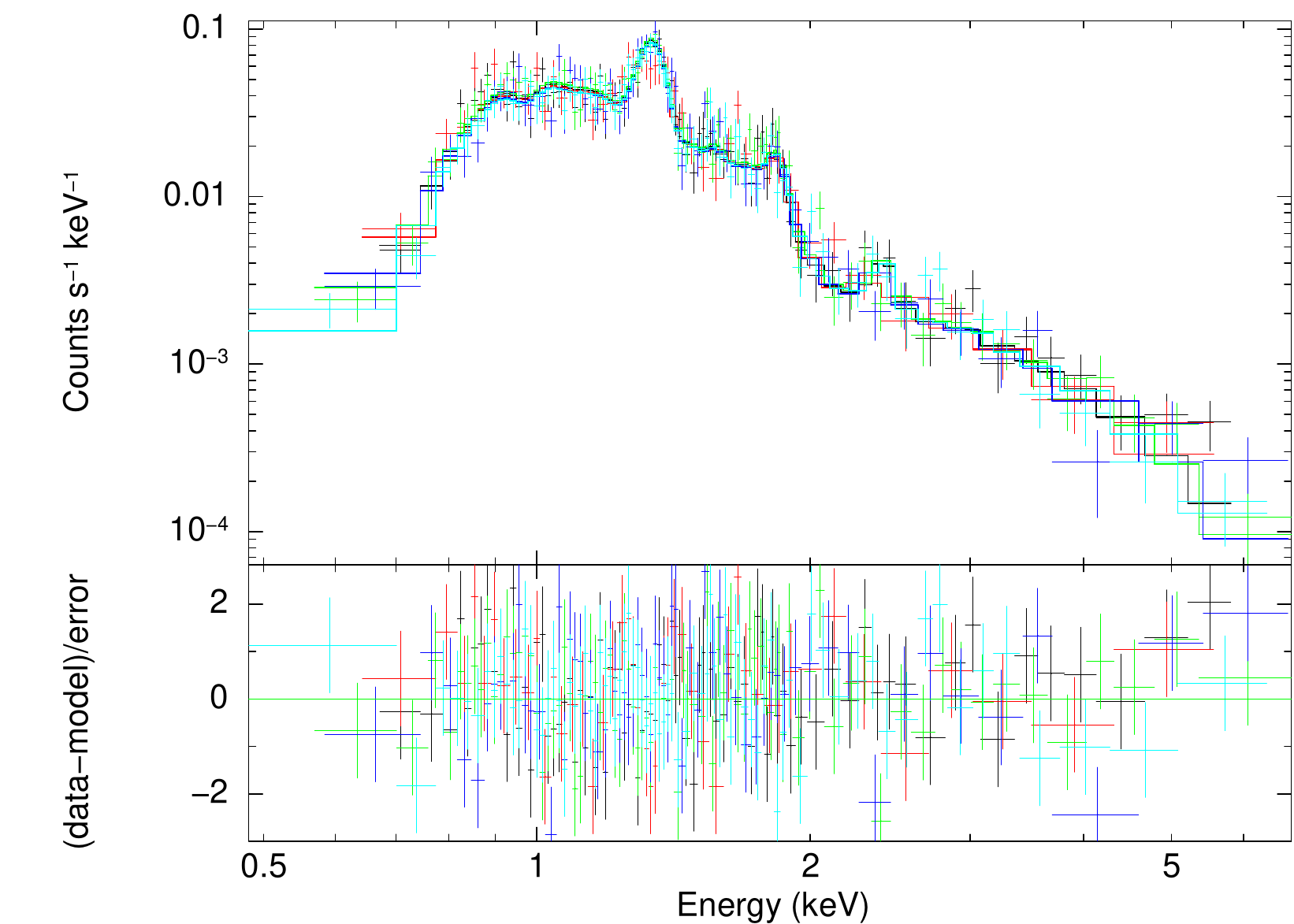}
		\label{fig:ejecta3spec}}
    \\
    \subfloat[ejecta4]{\includegraphics[height=0.48\columnwidth,angle=-90]{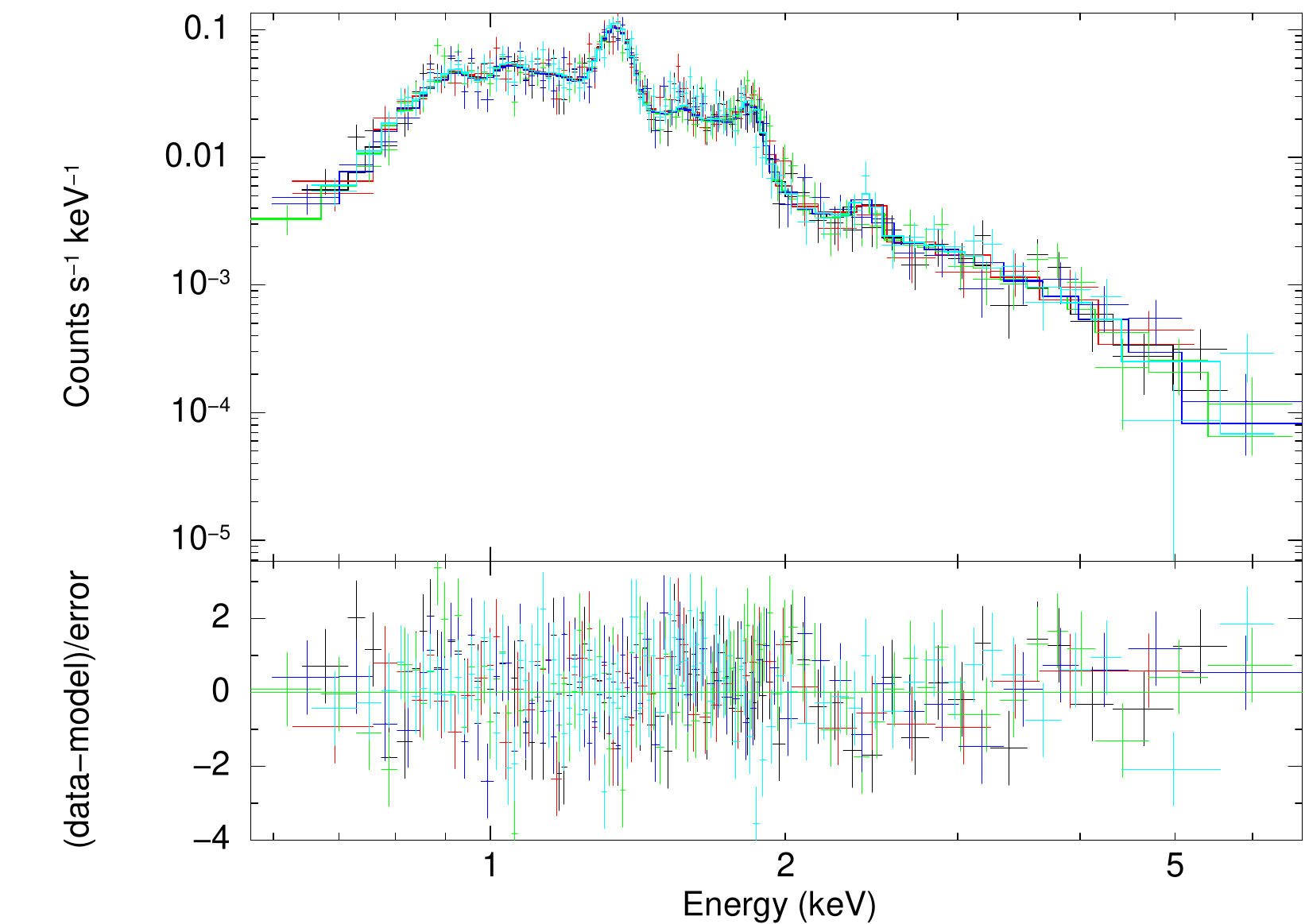}
		\label{fig:ejecta4spec}}
	\subfloat[clump]{\includegraphics[height=0.48\columnwidth,angle=-90]{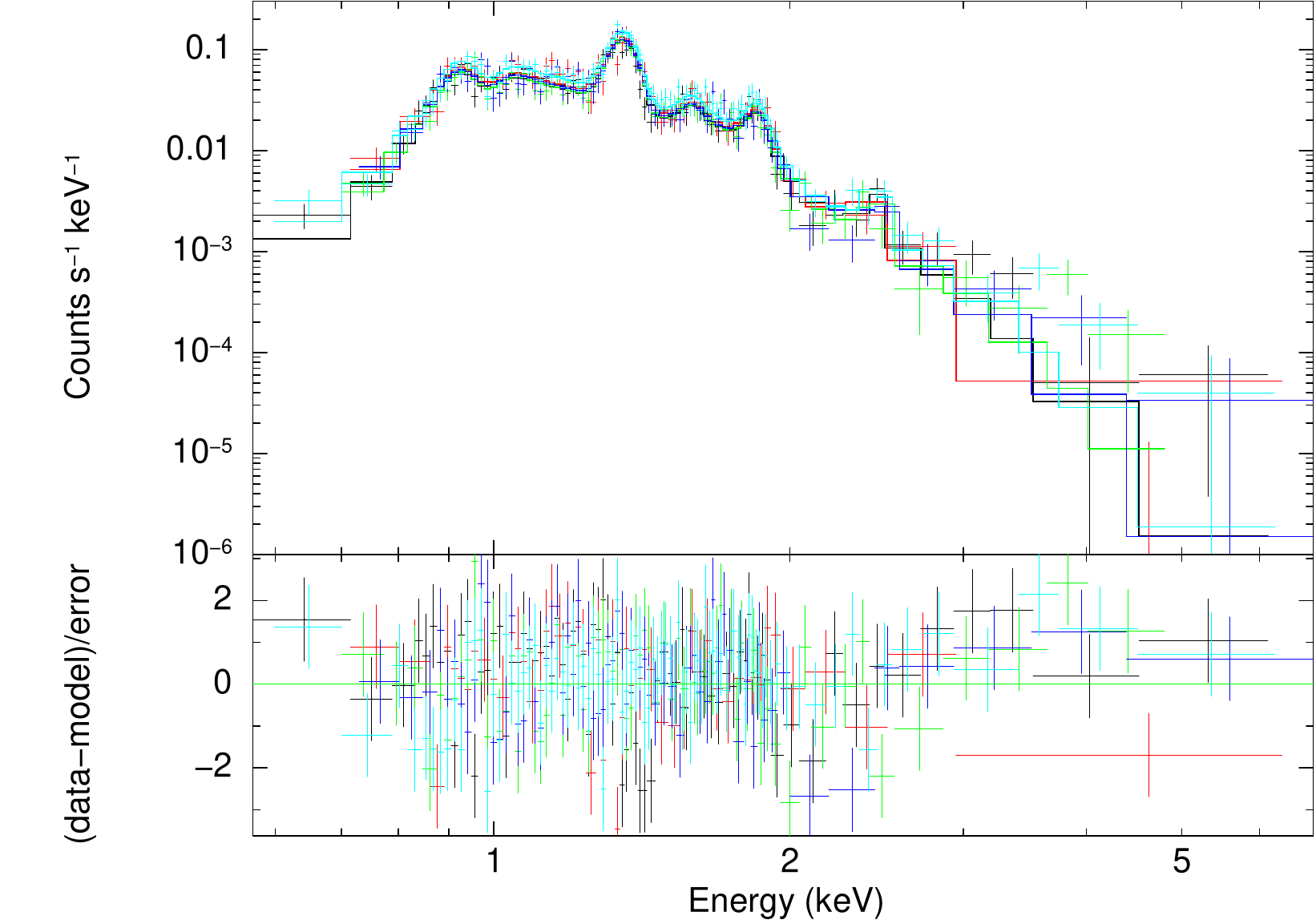}
		\label{fig:ejecta5spec}}
	\caption{\textit{Chandra} spectra of the regions ``ejecta2'' ``ejecta3'' ``ejecta4'', and ``clump'', as listed in Table \ref{tab:clump} and Table \ref{tab:ejecta}. 
    Black, red, green, blue, and cyan correspond to ObsIDs 19299, 19300, 20910, 20932, and 20933, respectively.
    All spectra are fitted simultaneously with the model \textit{TBabs}(\textit{vashift} $\times$ \textit{vpshock}).}
	\label{fig:chandraspec2}
\end{figure}

\begin{table*}[ht!]
\centering
\caption{Best-fit results and 90\% uncertainties for all jet regions. $N_{H}$ values free to vary.}
\label{tab:alljetregionfit_freenH}
\begin{tabular}{llcccc}
\hline\hline
Jet & Region & $N_{H}$ ($10^{22}$ cm$^{-2}$) & norm (10$^{-3}$ cm$^{-5}$) & Photon Index & $\chi^2_\nu$(d.o.f.)\\
\hline
South & Region 0 &  $1.28^{+0.04}_{-0.04}$ &$1.67^{+0.06}_{-0.06}$ & $1.69^{+0.03}_{-0.03}$ & 1.77(1179)\\
      & Region 1 & $1.24^{+0.05}_{-0.04}$& $1.03^{+0.05}_{-0.05}$ & $1.73^{+0.04}_{-0.04}$ & 1.17(974)\\
      & Region 2 & $1.35^{+0.05}_{-0.04}$& $1.32^{+0.06}_{-0.06}$ & $1.88^{+0.04}_{-0.04}$ & 1.24(982)\\
      & Region 3 & $1.48^{+0.05}_{-0.05}$& $1.35^{+0.07}_{-0.07}$ & $1.88^{+0.04}_{-0.04}$ & 1.27(982)\\
      & Region 4 & $1.56^{+0.07}_{-0.07}$& $1.32^{+0.09}_{-0.08}$ & $1.82^{+0.05}_{-0.05}$ & 0.99(1047)\\
      & Region 5 & $1.20^{+0.07}_{-0.07}$ & $1.02^{+0.08}_{-0.07}$ & $1.90^{+0.06}_{-0.06}$ & 1.11(946)\\
\hline
North & Region 0 & $1.43^{+0.05}_{-0.05}$& $1.02^{+0.05}_{-0.05}$ & $2.09^{+0.04}_{-0.04}$ & 1.65(1343)\\
      & Region 1 & $1.38^{+0.05}_{-0.05}$& $0.64^{+0.04}_{-0.03}$ & $2.08^{+0.05}_{-0.05}$ & 1.27(988)\\
      & Region 2 & $1.39^{+0.05}_{-0.04}$& $0.88^{+0.05}_{-0.04}$ & $2.27^{+0.04}_{-0.04}$ & 1.18(1082)\\
      & Region 3 & $1.40^{+0.05}_{-0.05}$& $1.06^{+0.05}_{-0.05}$ & $2.19^{+0.04}_{-0.04}$ & 1.33(1293)\\
      & Region 4 & $0.94^{+0.03}_{-0.03}$ & $1.02^{+0.04}_{-0.04}$ & $2.12^{+0.03}_{-0.03}$ & 1.10(1458)\\
      & Region 5 & $1.13^{+0.06}_{-0.06}$& $0.51^{+0.03}_{-0.03}$ & $2.12^{+0.05}_{-0.05}$ & 1.62(928)\\
\hline
\end{tabular}
\end{table*}

\clearpage
\bibliography{main}{}

@ARTICLE{2012AdSpR..49.1313F,
       author = {{Ferrand}, Gilles and {Safi-Harb}, Samar},
        title = "{A census of high-energy observations of Galactic supernova remnants}",
      journal = {Advances in Space Research},
         year = 2012,
        month = may,
       volume = {49},
       number = {9},
        pages = {1313-1319},
          doi = {10.1016/j.asr.2012.02.004},
archivePrefix = {arXiv},
       eprint = {1202.0245},
 primaryClass = {astro-ph.HE},
       adsurl = {https://ui.adsabs.harvard.edu/abs/2012AdSpR..49.1313F}
}

@BOOK{2017hsn..book.....A,
       author = {{Alsabti}, Athem W. and {Murdin}, Paul},
        title = "{Handbook of Supernovae}",
         year = 2017,
          doi = {10.1007/978-3-319-21846-5},
       adsurl = {https://ui.adsabs.harvard.edu/abs/2017hsn..book.....A}
}

@ARTICLE{2018ApJ...856...63F,
       author = {{Fryer}, Chris L. and {Andrews}, Sydney and {Even}, Wesley and {Heger}, Alex and {Safi-Harb}, Samar},
        title = "{Parameterizing the Supernova Engine and Its Effect on Remnants and Basic Yields}",
      journal = {\apj},
         year = 2018,
        month = mar,
       volume = {856},
       number = {1},
          eid = {63},
        pages = {63},
          doi = {10.3847/1538-4357/aaaf6f},
archivePrefix = {arXiv},
       eprint = {1712.03415},
 primaryClass = {astro-ph.HE},
       adsurl = {https://ui.adsabs.harvard.edu/abs/2018ApJ...856...63F}
}

@ARTICLE{2026ApJ...997..211T,
       author = {{Treyturik}, Cole and {Braun}, Chelsea and {Safi-Harb}, Samar and {Fryer}, Christopher L. and {Ferrand}, Gilles},
        title = "{Revisiting the Supernova Engines in the 3C 397 and W49B Supernova Remnants}",
      journal = {\apj},
         year = 2026,
        month = feb,
       volume = {997},
       number = {2},
          eid = {211},
        pages = {211},
          doi = {10.3847/1538-4357/ae2c50},
archivePrefix = {arXiv},
       eprint = {2512.01176},
 primaryClass = {astro-ph.HE},
       adsurl = {https://ui.adsabs.harvard.edu/abs/2026ApJ...997..211T}
}

@ARTICLE{2026ApJ..1003L...2B,
       author = {{Boccioli}, L. and {Roberti}, L. and {Fryer}, C. and {Safi-Harb}, S. and {Jones}, S. and {Pignatari}, M.},
        title = "{Production of Heavy {\ensuremath{\alpha}}-Elements and $^{44}$Ti in Cas A: Comparison to Abundances from 1D Core-collapse Supernova Models and Evidence for Carbon─Oxygen Shell Mergers}",
      journal = {\apjl},
         year = 2026,
        month = may,
       volume = {1003},
       number = {1},
          eid = {L2},
        pages = {L2},
          doi = {10.3847/2041-8213/ae626d},
archivePrefix = {arXiv},
       eprint = {2603.24758},
 primaryClass = {astro-ph.HE},
       adsurl = {https://ui.adsabs.harvard.edu/abs/2026ApJ..1003L...2B}
}

@article{Borkowski_2020,
doi = {10.3847/2041-8213/ab91c0},
url = {https://doi.org/10.3847/2041-8213/ab91c0},
year = {2020},
month = {may},
publisher = {The American Astronomical Society},
volume = {895},
number = {2},
pages = {L32},
author = {Borkowski, Kazimierz J. and Reynolds, Stephen P. and Miltich, William},
title = {Fast Blast Wave and Ejecta in the Young Core-collapse Supernova Remnant MSH 15-52/RCW 89},
journal = {The Astrophysical Journal Letters}
}

@article{Yatsu_2005,
doi = {10.1086/432590},
url = {https://doi.org/10.1086/432590},
year = {2005},
month = {sep},
publisher = {},
volume = {631},
number = {1},
pages = {312},
author = {Yatsu, Y. and Kawai, N. and Kataoka, J. and Kotani, T. and Tamura, K. and Brinkmann, W.},
title = {Chandra Observation of the Interaction between the Hot Plasma Nebula RCW 89 and the Pulsar Jet of PSR B1509–58},
journal = {The Astrophysical Journal}
}

@article{Gaensler_1999,
    author = {Gaensler, B. M. and Brazier, K. T. S. and Manchester, R. N. and Johnston, S. and Green, A. J.},
    title = {SNR G320.4–01.2 and PSR B1509–58: new radio observations of a complex interacting system},
    journal = {Monthly Notices of the Royal Astronomical Society},
    volume = {305},
    number = {3},
    pages = {724-736},
    year = {1999},
    month = {05},
    issn = {0035-8711},
    doi = {10.1046/j.1365-8711.1999.02500.x},
    url = {https://doi.org/10.1046/j.1365-8711.1999.02500.x},
    eprint = {https://academic.oup.com/mnras/article-pdf/305/3/724/3059292/305-3-724.pdf},
}

@ARTICLE{Trussoni_1996,
       author = {{Trussoni}, E. and {Massaglia}, S. and {Caucino}, S. and {Brinkmann}, W. and {Aschenbach}, B.},
        title = "{ROSAT PSPC observations of the supernova remnant MSH 15-52.}",
      journal = {\aap},
         year = 1996,
        month = feb,
       volume = {306},
        pages = {581},
       adsurl = {https://ui.adsabs.harvard.edu/abs/1996A\&A...306..581T}
}

@ARTICLE{Brazier_1997MNRAS,
       author = {{Brazier}, K.~T.~S. and {Becker}, W.},
        title = "{High-resolution X-ray imaging of the supernova remnant MSH 15-52}",
      journal = {\mnras},
         year = 1997,
        month = jan,
       volume = {284},
       number = {2},
        pages = {335-340},
          doi = {10.1093/mnras/284.2.335},
       adsurl = {https://ui.adsabs.harvard.edu/abs/1997MNRAS.284..335B}
}

@ARTICLE{Manchester_1987,
       author = {{Manchester}, R.~N.},
        title = "{The radio structure of supernova remnants.}",
      journal = {\aap},
         year = 1987,
        month = jan,
       volume = {171},
        pages = {205-215},
       adsurl = {https://ui.adsabs.harvard.edu/abs/1987A\&A...171..205M}
}

@ARTICLE{Seward_1983,
       author = {{Seward}, F.~D. and {Harnden}, Jr., F.~R. and {Murdin}, P. and {Clark}, D.~H.},
        title = "{MSH 15-52 : a supernova remnant containing two compact X-ray sources.}",
      journal = {\apj},
         year = 1983,
        month = apr,
       volume = {267},
        pages = {698-710},
          doi = {10.1086/160907},
       adsurl = {https://ui.adsabs.harvard.edu/abs/1983ApJ...267..698S}
}

@ARTICLE{Sukhbold2016,
       author = {{Sukhbold}, Tuguldur and {Ertl}, T. and {Woosley}, S.~E. and {Brown}, Justin M. and {Janka}, H. -T.},
        title = "{Core-collapse Supernovae from 9 to 120 Solar Masses Based on Neutrino-powered Explosions}",
      journal = {\apj},
         year = 2016,
        month = apr,
       volume = {821},
       number = {1},
          eid = {38},
        pages = {38},
          doi = {10.3847/0004-637X/821/1/38},
archivePrefix = {arXiv},
       eprint = {1510.04643},
 primaryClass = {astro-ph.HE},
       adsurl = {https://ui.adsabs.harvard.edu/abs/2016ApJ...821...38S}
}

@article{Zhang_2025,
doi = {10.3847/1538-4357/adf333},
url = {https://doi.org/10.3847/1538-4357/adf333},
year = {2025},
month = {aug},
publisher = {The American Astronomical Society},
volume = {989},
number = {2},
pages = {221},
author = {Zhang, S. and Ng, C.-Y. and Bucciantini, N.},
title = {High-resolution Radio Study of Pulsar Wind Nebula MSH 15–52 and Supernova Remnant RCW 89},
journal = {The Astrophysical Journal}
}

@ARTICLE{kaspi_1994,
       author = {{Kaspi}, V.~M. and {Manchester}, R.~N. and {Siegman}, B. and {Johnston}, S. and {Lyne}, A.~G.},
        title = "{On the Spin-down of PSR B1509-58}",
      journal = {\apjl},
         year = 1994,
        month = feb,
       volume = {422},
        pages = {L83},
          doi = {10.1086/187218},
       adsurl = {https://ui.adsabs.harvard.edu/abs/1994ApJ...422L..83K}
}

@ARTICLE{Ye_2025,
       author = {{Ye}, W.~T. and {Zhang}, P.~F. and {Ge}, M.~Y. and {Qi}, L.~Q. and {Zheng}, S.~J. and {Tuo}, Y.~L. and {Han}, D.~W. and {Peng}, H.~L. and {Xiao}, Y.~X. and {Shui}, Q.~C. and {Lu}, F.~J.},
        title = "{Constraints on Emission Models by Broadband Observations of Soft Gamma-Ray Pulsar PSR B1509{\textendash}58}",
      journal = {\apj},
         year = 2025,
        month = nov,
       volume = {993},
       number = {1},
          eid = {108},
        pages = {108},
          doi = {10.3847/1538-4357/ae0303},
       adsurl = {https://ui.adsabs.harvard.edu/abs/2025ApJ...993..108Y}
}

@article{Safi-Harb_2022,
doi = {10.3847/1538-4357/ac7c05},
url = {https://doi.org/10.3847/1538-4357/ac7c05},
year = {2022},
month = {aug},
publisher = {The American Astronomical Society},
volume = {935},
number = {2},
pages = {163},
author = {Safi-Harb, Samar and Mac Intyre, Brydyn and Zhang, Shuo and Pope, Isaac and Zhang, Shuhan and Saffold, Nathan and Mori, Kaya and Gotthelf, Eric V. and Aharonian, Felix and Band, Matthew and Braun, Chelsea and Fang, Ke and Hailey, Charles and Nynka, Melania and Rho, Chang D.},
title = {Hard X-Ray Emission from the Eastern Jet of SS 433 Powering the W50 “Manatee” Nebula: Evidence for Particle Reacceleration},
journal = {The Astrophysical Journal}
}

@ARTICLE{Aharonian_2005,
       author = {{Aharonian}, F. and {Akhperjanian}, A.~G. and {Aye}, K.-M. and {Bazer-Bachi}, A.~R. and {Beilicke}, M. and {Benbow}, W. and {Berge}, D. and {Berghaus}, P. and {Bernl{\"o}hr}, K. and {Boisson}, C. and {Bolz}, O. and {Braun}, I. and {Breitling}, F. and {Brown}, A.~M. and {Bussons Gordo}, J. and {Chadwick}, P.~M. and {Chounet}, L.-M. and {Cornils}, R. and {Costamante}, L. and {Degrange}, B. and {Djannati-Ata{\"\i}}, A. and {O'C. Drury}, L. and {Dubus}, G. and {Emmanoulopoulos}, D. and {Espigat}, P. and {Feinstein}, F. and {Fleury}, P. and {Fontaine}, G. and {Fuchs}, Y. and {Funk}, S. and {Gallant}, Y.~A. and {Giebels}, B. and {Gillessen}, S. and {Glicenstein}, J.~F. and {Goret}, P. and {Hadjichristidis}, C. and {Hauser}, M. and {Heinzelmann}, G. and {Henri}, G. and {Hermann}, G. and {Hinton}, J.~A. and {Hofmann}, W. and {Holleran}, M. and {Horns}, D. and {de Jager}, O.~C. and {Kh{\'e}lifi}, B. and {Komin}, Nu. and {Konopelko}, A. and {Latham}, I.~J. and {Le Gallou}, R. and {Lemi{\`e}re}, A. and {Lemoine-Goumard}, M. and {Leroy}, N. and {Lohse}, T. and {Martineau-Huynh}, O. and {Marcowith}, A. and {Masterson}, C. and {McComb}, T.~J.~L. and {de Naurois}, M. and {Nolan}, S.~J. and {Noutsos}, A. and {Orford}, K.~J. and {Osborne}, J.~L. and {Ouchrif}, M. and {Panter}, M. and {Pelletier}, G. and {Pita}, S. and {P{\"u}hlhofer}, G. and {Punch}, M. and {Raubenheimer}, B.~C. and {Raue}, M. and {Raux}, J. and {Rayner}, S.~M. and {Redondo}, I. and {Reimer}, A. and {Reimer}, O. and {Ripken}, J. and {Rob}, L. and {Rolland}, L. and {Rowell}, G. and {Sahakian}, V. and {Saug{\'e}}, L. and {Schlenker}, S. and {Schlickeiser}, R. and {Schuster}, C. and {Schwanke}, U. and {Siewert}, M. and {Sol}, H. and {Steenkamp}, R. and {Stegmann}, C. and {Tavernet}, J.-P. and {Terrier}, R. and {Th{\'e}oret}, C.~G. and {Tluczykont}, M. and {Vasileiadis}, G. and {Venter}, C. and {Vincent}, P. and {V{\"o}lk}, H.~J. and {Wagner}, S.~J.},
        title = "{Discovery of extended VHE gamma-ray emission from the asymmetric pulsar wind nebula in MSH 15-52 with HESS}",
      journal = {\aap},
         year = 2005,
        month = may,
       volume = {435},
       number = {1},
        pages = {L17-L20},
          doi = {10.1051/0004-6361:200500105},
archivePrefix = {arXiv},
       eprint = {astro-ph/0504120},
 primaryClass = {astro-ph},
       adsurl = {https://ui.adsabs.harvard.edu/abs/2005A\&A...435L..17A}
}

@ARTICLE{2009_lopez,
       author = {{Lopez}, L.~A. and {Ramirez-Ruiz}, E. and {Badenes}, C. and {Huppenkothen}, D. and {Jeltema}, T.~E. and {Pooley}, D.~A.},
        title = "{Typing Supernova Remnants Using X-Ray Line Emission Morphologies}",
      journal = {\apjl},
         year = 2009,
        month = nov,
       volume = {706},
       number = {1},
        pages = {L106-L109},
          doi = {10.1088/0004-637X/706/1/L106},
archivePrefix = {arXiv},
       eprint = {0910.3208},
 primaryClass = {astro-ph.HE},
       adsurl = {https://ui.adsabs.harvard.edu/abs/2009ApJ...706L.106L}
}

@ARTICLE{2014_yamaguchi,
       author = {{Yamaguchi}, Hiroya and {Badenes}, Carles and {Petre}, Robert and {Nakano}, Toshio and {Castro}, Daniel and {Enoto}, Teruaki and {Hiraga}, Junko S. and {Hughes}, John P. and {Maeda}, Yoshitomo and {Nobukawa}, Masayoshi and {Safi-Harb}, Samar and {Slane}, Patrick O. and {Smith}, Randall K. and {Uchida}, Hiroyuki},
        title = "{Discriminating the Progenitor Type of Supernova Remnants with Iron K-shell Emission}",
      journal = {\apjl},
         year = 2014,
        month = apr,
       volume = {785},
       number = {2},
          eid = {L27},
        pages = {L27},
          doi = {10.1088/2041-8205/785/2/L27},
archivePrefix = {arXiv},
       eprint = {1403.5154},
 primaryClass = {astro-ph.HE},
       adsurl = {https://ui.adsabs.harvard.edu/abs/2014ApJ...785L..27Y}
}

@ARTICLE{2023_braun,
       author = {{Braun}, C. and {Safi-Harb}, S. and {Fryer}, C.~L. and {Zhou}, P.},
        title = "{Progenitors and explosion properties of supernova remnants hosting central compact objects: II. A global systematic study with a comparison to nucleosynthesis models}",
      journal = {\mnras},
         year = 2023,
        month = nov,
       volume = {525},
       number = {4},
        pages = {6257-6284},
          doi = {10.1093/mnras/stad2592},
archivePrefix = {arXiv},
       eprint = {2308.13693},
 primaryClass = {astro-ph.HE},
       adsurl = {https://ui.adsabs.harvard.edu/abs/2023MNRAS.525.6257B}
}

@ARTICLE{Zhou_2019,
       author = {{Zhou}, Ping and {Vink}, Jacco and {Safi-Harb}, Samar and {Miceli}, Marco},
        title = "{Spatially resolved X-ray study of supernova remnants that host magnetars: Implication of their fossil field origin}",
      journal = {\aap},
         year = 2019,
        month = sep,
       volume = {629},
          eid = {A51},
        pages = {A51},
          doi = {10.1051/0004-6361/201936002},
archivePrefix = {arXiv},
       eprint = {1909.01922},
 primaryClass = {astro-ph.HE},
       adsurl = {https://ui.adsabs.harvard.edu/abs/2019A&A...629A..51Z}
}

@ARTICLE{1982_seward,
       author = {{Seward}, F.~D. and {Harnden}, Jr., F.~R.},
        title = "{A new, fast X-ray pulsar in the supernova remnant MSH 15-52.}",
      journal = {\apjl},
         year = 1982,
        month = may,
       volume = {256},
        pages = {L45-L47},
          doi = {10.1086/183793},
       adsurl = {https://ui.adsabs.harvard.edu/abs/1982ApJ...256L..45S}
}

@article{1961_Mills,
    author = {Mills, BY and Slee, OB and Hill, ER},
    title = {A Catalogue of Radio Sources Between Declinations ?50° and ?80°},
    journal = {Australian Journal of Physics},
    volume = {14},
    number = {4},
    pages = {497-507},
    year = {1961},
    month = {12},
    issn = {0004-9506},
    doi = {10.1071/PH610497},
    url = {https://doi.org/10.1071/PH610497},
    eprint = {https://connectsci.au/ph/article-pdf/14/4/497/1347542/ph610497.pdf},
}

@ARTICLE{2002_gaensler,
       author = {{Gaensler}, B.~M. and {Arons}, J. and {Kaspi}, V.~M. and {Pivovaroff}, M.~J. and {Kawai}, N. and {Tamura}, K.},
        title = "{Chandra Imaging of the X-Ray Nebula Powered by Pulsar B1509-58}",
      journal = {\apj},
         year = 2002,
        month = apr,
       volume = {569},
       number = {2},
        pages = {878-893},
          doi = {10.1086/339354},
archivePrefix = {arXiv},
       eprint = {astro-ph/0110454},
 primaryClass = {astro-ph},
       adsurl = {https://ui.adsabs.harvard.edu/abs/2002ApJ...569..878G}
}

@ARTICLE{2025_Egg,
       author = {{Egg}, Katharina and {Mitchell}, Alison M.~W.},
        title = "{Joint eROSITA and H.E.S.S. analysis of MSH 15-52 using Gammapy}",
      journal = {arXiv e-prints},
         year = 2025,
        month = oct,
          eid = {arXiv:2510.04857},
        pages = {arXiv:2510.04857},
          doi = {10.48550/arXiv.2510.04857},
archivePrefix = {arXiv},
       eprint = {2510.04857},
 primaryClass = {astro-ph.HE},
       adsurl = {https://ui.adsabs.harvard.edu/abs/2025arXiv251004857E}
}

@inproceedings{2006Fruscione,
  author    = {Fruscione, A. and McDowell, J. C. and Allen, G. E. and others},
  title     = {CIAO: Chandra's data analysis system},
  booktitle = {Society of Photo-Optical Instrumentation Engineers (SPIE) Conference Series},
  volume    = {6270},
  year      = {2006},
  doi       = {10.1117/12.671760}
}

@software{2014SAS,
       author = {{SAS development team}},
        title = "{SAS: Science Analysis System for XMM-Newton observatory}",
 howpublished = {Astrophysics Source Code Library, record ascl:1404.004},
         year = 2014,
        month = apr,
          eid = {ascl:1404.004},
archivePrefix = {ascl},
       eprint = {1404.004},
       adsurl = {https://ui.adsabs.harvard.edu/abs/2014ascl.soft04004S}
}

@article{2010_schock,
	author = {{Schöck}, F. M. and {Büsching}, I. and {de Jager}, O. C. and {Eger}, P. and {Vorster}, M. J.},
	title = {Spatially resolved XMM-Newton analysis and a model   of the nonthermal emission of MSH 15–52},
	DOI= "10.1051/0004-6361/201014151",
	url= "https://doi.org/10.1051/0004-6361/201014151",
	journal = {A\&A},
	year = 2010,
	volume = 515,
	pages = "A109",
	month = "",
}

@INPROCEEDINGS{2003Ds9_Joye,
       author = {{Joye}, W.~A. and {Mandel}, E.},
        title = "{New Features of SAOImage DS9}",
    booktitle = {Astronomical Data Analysis Software and Systems XII},
         year = 2003,
       editor = {{Payne}, H.~E. and {Jedrzejewski}, R.~I. and {Hook}, R.~N.},
       series = {Astronomical Society of the Pacific Conference Series},
       volume = {295},
        month = jan,
        pages = {489},
       adsurl = {https://ui.adsabs.harvard.edu/abs/2003ASPC..295..489J}
}

@INPROCEEDINGS{1996xspec_Arnaud,
       author = {{Arnaud}, K.~A.},
        title = "{XSPEC: The First Ten Years}",
    booktitle = {Astronomical Data Analysis Software and Systems V},
         year = 1996,
       editor = {{Jacoby}, George H. and {Barnes}, Jeannette},
       series = {Astronomical Society of the Pacific Conference Series},
       volume = {101},
        month = jan,
        pages = {17},
       adsurl = {https://ui.adsabs.harvard.edu/abs/1996ASPC..101...17A}
}

@ARTICLE{Contbin,
       author = {{Sanders}, J.~S.},
        title = "{Contour binning: a new technique for spatially resolved X-ray spectroscopy applied to Cassiopeia A}",
      journal = {\mnras},
         year = 2006,
        month = sep,
       volume = {371},
       number = {2},
        pages = {829-842},
          doi = {10.1111/j.1365-2966.2006.10716.x},
archivePrefix = {arXiv},
       eprint = {astro-ph/0606528},
 primaryClass = {astro-ph},
       adsurl = {https://ui.adsabs.harvard.edu/abs/2006MNRAS.371..829S}
}

@article{vanderSwaluw_2004,
	author = {{van der Swaluw, E.} and {Downes, T. P.} and {Keegan, R.}},
	title = {An evolutionary model for pulsar-driven supernova remnants - A hydrodynamical model},
	DOI= "10.1051/0004-6361:20035700",
	url= "https://doi.org/10.1051/0004-6361:20035700",
	journal = {A\&A},
	year = 2004,
	volume = 420,
	number = 3,
	pages = "937-944",
}

@article{An_2014,
doi = {10.1088/0004-637X/793/2/90},
url = {https://doi.org/10.1088/0004-637X/793/2/90},
year = {2014},
month = {sep},
publisher = {The American Astronomical Society},
volume = {793},
number = {2},
pages = {90},
author = {An, Hongjun and Madsen, Kristin K. and Reynolds, Stephen P. and Kaspi, Victoria M. and Harrison, Fiona A. and Boggs, Steven E. and Christensen, Finn E. and Craig, William W. and Fryer, Chris L. and Grefenstette, Brian W. and Hailey, Charles J. and Mori, Kaya and Stern, Daniel and Zhang, William W.},
title = {HIGH-ENERGY X-RAY IMAGING OF THE PULSAR WIND NEBULA MSH 15−52: CONSTRAINTS ON PARTICLE ACCELERATION AND TRANSPORT},
journal = {The Astrophysical Journal}
}

@article{Romani_2023,
doi = {10.3847/1538-4357/acfa02},
url = {https://doi.org/10.3847/1538-4357/acfa02},
year = {2023},
month = {oct},
publisher = {The American Astronomical Society},
volume = {957},
number = {1},
pages = {23},
author = {Romani, Roger W. and Wong, Josephine and Di Lalla, Niccoló and Omodei, Nicola and Xie, Fei and Ng, C.-Y. and Ferrazzoli, Riccardo and Di Marco, Alessandro and Bucciantini, Niccoló and Pilia, Maura and Slane, Patrick and Weisskopf, Martin C. and Johnston, Simon and Burgay, Marta and Wei, Deng and Yang, Yi-Jung and Zhang, Shumeng and Antonelli, Lucio A. and Bachetti, Matteo and Baldini, Luca and Baumgartner, Wayne H. and Bellazzini, Ronaldo and Bianchi, Stefano and Bongiorno, Stephen D. and Bonino, Raffaella and Brez, Alessandro and Capitanio, Fiamma and Castellano, Simone and Cavazzuti, Elisabetta and Chen, Chien-Ting and Cibrario, Nicoló and Ciprini, Stefano and Costa, Enrico and De Rosa, Alessandra and Del Monte, Ettore and Di Gesu, Laura and Donnarumma, Immacolata and Doroshenko, Victor and Dovčiak, Michal and Ehlert, Steven R. and Enoto, Teruaki and Evangelista, Yuri and Fabiani, Sergio and Garcia, Javier A. and Gunji, Shuichi and Hayashida, Kiyoshi and Heyl, Jeremy and Iwakiri, Wataru and Liodakis, Ioannis and Kaaret, Philip and Karas, Vladimir and Kim, Dawoon E. and Kitaguchi, Takao and Kolodziejczak, Jeffery J. and Krawczynski, Henric and La Monaca, Fabio and Latronico, Luca and Madejski, Grzegorz and Maldera, Simone and Manfreda, Alberto and Marin, Frédéric and Marinucci, Andrea and Marscher, Alan P. and Marshall, Herman L. and Massaro, Francesco and Matt, Giorgio and Middei, Riccardo and Mitsuishi, Ikuyuki and Mizuno, Tsunefumi and Muleri, Fabio and Negro, Michela and O’Dell, Stephen L. and Oppedisano, Chiara and Pacciani, Luigi and Papitto, Alessandro and Pavlov, George G. and Perri, Matteo and Pesce-Rollins, Melissa and Petrucci, Pierre-Olivier and Possenti, Andrea and Poutanen, Juri and Puccetti, Simonetta and Ramsey, Brian D. and Rankin, John and Ratheesh, Ajay and Roberts, Oliver J. and Sgró, Carmelo and Soffitta, Paolo and Spandre, Gloria and Swartz, Douglas A. and Tamagawa, Toru and Tavecchio, Fabrizio and Taverna, Roberto and Tawara, Yuzuru and Tennant, Allyn F. and Thomas, Nicholas E. and Tombesi, Francesco and Trois, Alessio and Tsygankov, Sergey and Turolla, Roberto and Vink, Jacco and Wu, Kinwah and Zane, Silvia},
title = {The Polarized Cosmic Hand: IXPE Observations of PSR B1509−58/MSH 15−52},
journal = {The Astrophysical Journal}
}

@article{DeLaney_2006,
   title={Time Variability in the X‐Ray Nebula Powered by Pulsar B1509−58},
   volume={640},
   ISSN={1538-4357},
   url={http://dx.doi.org/10.1086/500189},
   DOI={10.1086/500189},
   number={2},
   journal={The Astrophysical Journal},
   publisher={American Astronomical Society},
   author={DeLaney, T. and Gaensler, B. M. and Arons, J. and Pivovaroff, M. J.},
   year={2006},
   month=apr, pages={929–940} }

@ARTICLE{2020ApJ...890...35A,
       author = {{Andrews}, S. and {Fryer}, C. and {Even}, W. and {Jones}, S. and {Pignatari}, M.},
        title = "{The Nucleosynthetic Yields of Core-collapse Supernovae: Prospects for the Next Generation of Gamma-Ray Astronomy}",
      journal = {\apj},
         year = 2020,
        month = feb,
       volume = {890},
       number = {1},
          eid = {35},
        pages = {35},
          doi = {10.3847/1538-4357/ab64f8},
archivePrefix = {arXiv},
       eprint = {1912.10542},
 primaryClass = {astro-ph.HE},
       adsurl = {https://ui.adsabs.harvard.edu/abs/2020ApJ...890...35A}
}

@INPROCEEDINGS{esass_brunner_2018,
       author = {{Brunner}, H. and {Boller}, T. and {Coutinho}, D. and {Dauser}, T. and {Dennerl}, K. and {Dwelly}, T. and {Freyberg}, M. and {F{\"u}rmetz}, M. and {Georgakakis}, A. and {Grossberger}, C. and {Kreykenbohm}, I. and {Lamer}, G. and {Meidinger}, N. and {M{\"u}ller}, S. and {Predehl}, P. and {Robrade}, J. and {Sanders}, J. and {Wilms}, J.},
        title = "{eROSITA ground operations}",
    booktitle = {Space Telescopes and Instrumentation 2018: Ultraviolet to Gamma Ray},
         year = 2018,
       editor = {{den Herder}, Jan-Willem A. and {Nikzad}, Shouleh and {Nakazawa}, Kazuhiro},
       series = {Society of Photo-Optical Instrumentation Engineers (SPIE) Conference Series},
       volume = {10699},
        month = jul,
          eid = {106995G},
        pages = {106995G},
          doi = {10.1117/12.2315138},
       adsurl = {https://ui.adsabs.harvard.edu/abs/2018SPIE10699E..5GB}
}

@ARTICLE{2021_Predehl_esass,
       author = {{Predehl}, P. and {Andritschke}, R. and {Arefiev}, V. and {Babyshkin}, V. and {Batanov}, O. and {Becker}, W. and {B{\"o}hringer}, H. and {Bogomolov}, A. and {Boller}, T. and {Borm}, K. and {Bornemann}, W. and {Br{\"a}uninger}, H. and {Br{\"u}ggen}, M. and {Brunner}, H. and {Brusa}, M. and {Bulbul}, E. and {Buntov}, M. and {Burwitz}, V. and {Burkert}, W. and {Clerc}, N. and {Churazov}, E. and {Coutinho}, D. and {Dauser}, T. and {Dennerl}, K. and {Doroshenko}, V. and {Eder}, J. and {Emberger}, V. and {Eraerds}, T. and {Finoguenov}, A. and {Freyberg}, M. and {Friedrich}, P. and {Friedrich}, S. and {F{\"u}rmetz}, M. and {Georgakakis}, A. and {Gilfanov}, M. and {Granato}, S. and {Grossberger}, C. and {Gueguen}, A. and {Gureev}, P. and {Haberl}, F. and {H{\"a}lker}, O. and {Hartner}, G. and {Hasinger}, G. and {Huber}, H. and {Ji}, L. and {Kienlin}, A. v. and {Kink}, W. and {Korotkov}, F. and {Kreykenbohm}, I. and {Lamer}, G. and {Lomakin}, I. and {Lapshov}, I. and {Liu}, T. and {Maitra}, C. and {Meidinger}, N. and {Menz}, B. and {Merloni}, A. and {Mernik}, T. and {Mican}, B. and {Mohr}, J. and {M{\"u}ller}, S. and {Nandra}, K. and {Nazarov}, V. and {Pacaud}, F. and {Pavlinsky}, M. and {Perinati}, E. and {Pfeffermann}, E. and {Pietschner}, D. and {Ramos-Ceja}, M.~E. and {Rau}, A. and {Reiffers}, J. and {Reiprich}, T.~H. and {Robrade}, J. and {Salvato}, M. and {Sanders}, J. and {Santangelo}, A. and {Sasaki}, M. and {Scheuerle}, H. and {Schmid}, C. and {Schmitt}, J. and {Schwope}, A. and {Shirshakov}, A. and {Steinmetz}, M. and {Stewart}, I. and {Str{\"u}der}, L. and {Sunyaev}, R. and {Tenzer}, C. and {Tiedemann}, L. and {Tr{\"u}mper}, J. and {Voron}, V. and {Weber}, P. and {Wilms}, J. and {Yaroshenko}, V.},
        title = "{The eROSITA X-ray telescope on SRG}",
      journal = {\aap},
         year = 2021,
        month = mar,
       volume = {647},
          eid = {A1},
        pages = {A1},
          doi = {10.1051/0004-6361/202039313},
archivePrefix = {arXiv},
       eprint = {2010.03477},
 primaryClass = {astro-ph.HE},
       adsurl = {https://ui.adsabs.harvard.edu/abs/2021A&A...647A...1P}
}

@ARTICLE{Truelove_1999,
       author = {{Truelove}, J. Kelly and {McKee}, Christopher F.},
        title = "{Evolution of Nonradiative Supernova Remnants}",
      journal = {\apjs},
         year = 1999,
        month = feb,
       volume = {120},
       number = {2},
        pages = {299-326},
          doi = {10.1086/313176},
       adsurl = {https://ui.adsabs.harvard.edu/abs/1999ApJS..120..299T}
}

@BOOK{Sedov_1959,
       author = {{Sedov}, L.~I.},
        title = "{Similarity and Dimensional Methods in Mechanics}",
         year = 1959,
       adsurl = {https://ui.adsabs.harvard.edu/abs/1959sdmm.book.....S}
}

@ARTICLE{1967ZelDovich,
       author = {{Zel'Dovich}, Ya. B. and {Raizer}, Yu. P. and {Hayes}, Wallace D. and {Probstein}, Ronald F. and {Gill}, Stephen P.},
        title = "{Physics of Shock Waves and High-Temperature Hydrodynamic Phenomena, Vol. 1}",
      journal = {Journal of Applied Mechanics},
         year = 1967,
        month = jan,
       volume = {34},
       number = {4},
        pages = {1055},
          doi = {10.1115/1.3607836},
       adsurl = {https://ui.adsabs.harvard.edu/abs/1967JAM....34.1055Z}
}

@ARTICLE{2001Borkowski,
       author = {{Borkowski}, Kazimierz J. and {Lyerly}, William J. and {Reynolds}, Stephen P.},
        title = "{Supernova Remnants in the Sedov Expansion Phase: Thermal X-Ray Emission}",
      journal = {\apj},
         year = 2001,
        month = feb,
       volume = {548},
       number = {2},
        pages = {820-835},
          doi = {10.1086/319011},
archivePrefix = {arXiv},
       eprint = {astro-ph/0008066},
 primaryClass = {astro-ph},
       adsurl = {https://ui.adsabs.harvard.edu/abs/2001ApJ...548..820B}
}

@ARTICLE{1950Taylor,
       author = {{Taylor}, Geoffrey},
        title = "{The Formation of a Blast Wave by a Very Intense Explosion. I. Theoretical Discussion}",
      journal = {Proceedings of the Royal Society of London Series A},
         year = 1950,
        month = mar,
       volume = {201},
       number = {1065},
        pages = {159-174},
          doi = {10.1098/rspa.1950.0049},
       adsurl = {https://ui.adsabs.harvard.edu/abs/1950RSPSA.201..159T}
}

@ARTICLE{1950Taylor2,
       author = {{Taylor}, Geoffrey},
        title = "{The Formation of a Blast Wave by a Very Intense Explosion. II. The Atomic Explosion of 1945}",
      journal = {Proceedings of the Royal Society of London Series A},
         year = 1950,
        month = mar,
       volume = {201},
       number = {1065},
        pages = {175-186},
          doi = {10.1098/rspa.1950.0050},
       adsurl = {https://ui.adsabs.harvard.edu/abs/1950RSPSA.201..175T}
}

@ARTICLE{2025soker,
       author = {{Soker}, Noam},
        title = "{Attributing the Supernova Remnant RCW 89 to the Jittering Jets Explosion Mechanism}",
      journal = {\pasp},
         year = 2025,
        month = nov,
       volume = {137},
       number = {11},
          eid = {114201},
        pages = {114201},
          doi = {10.1088/1538-3873/ae1415},
archivePrefix = {arXiv},
       eprint = {2509.04723},
 primaryClass = {astro-ph.HE},
       adsurl = {https://ui.adsabs.harvard.edu/abs/2025PASP..137k4201S}
}

@article{Kolb_2017,
doi = {10.3847/1538-4357/aa75ce},
url = {https://doi.org/10.3847/1538-4357/aa75ce},
year = {2017},
month = {jul},
publisher = {The American Astronomical Society},
volume = {844},
number = {1},
pages = {1},
author = {Kolb, Christopher and Blondin, John and Slane, Patrick and Temim, Tea},
title = {Evolution of a Pulsar Wind Nebula within a Composite Supernova Remnant},
journal = {The Astrophysical Journal}
}

@article{Temim_2017,
doi = {10.3847/1538-4357/aa9d41},
url = {https://doi.org/10.3847/1538-4357/aa9d41},
year = {2017},
month = {dec},
publisher = {The American Astronomical Society},
volume = {851},
number = {2},
pages = {128},
author = {Temim, Tea and Slane, Patrick and Plucinsky, Paul P. and Gelfand, Joseph and Castro, Daniel and Kolb, Christopher},
title = {Proper Motion of the High-velocity Pulsar in SNR MSH 15-56},
journal = {The Astrophysical Journal}
}
\bibliographystyle{aasjournalv7}

\end{document}